\documentclass[aps,prl,reprint,superscriptaddress,amsmath,footinbib]{revtex4-2}
\usepackage[colorlinks=true,citecolor=blue,urlcolor=blue,linkcolor=blue]{hyperref}
\usepackage{graphicx}
\usepackage{placeins}
\usepackage{braket}
\usepackage{amssymb}
\usepackage{bbm}
\usepackage{xr}
\usepackage[normalem]{ulem}
\graphicspath{{../Figures/}}
\DeclareGraphicsExtensions{.pdf,.png}
\DeclareMathOperator{\Tr}{Tr}
\newcommand{\ketbra}[2]{\mathinner{|{#1}\rangle\langle{#2}|}}

\begin{document}
\title{Topological Burning Glass Effect}

\author{Simon K\"{o}rber}
\email[Email:\,]{skoerber@physik.uni-wuerzburg.de}
\affiliation{Institute of Theoretical Physics and Astrophysics, University of W\"{u}rzburg, Am Hubland, 97074 W\"{u}rzburg, Germany}

\author{Lorenzo Privitera}
\affiliation{Institute of Theoretical Physics and Astrophysics, University of W\"{u}rzburg, Am Hubland, 97074 W\"{u}rzburg, Germany}
\affiliation{Corporate Research, ABB Switzerland Ltd., Segelhofstrasse 1K, Baden-Dättwil 5405, Switzerland}

\author{Jan Carl Budich}
\affiliation{Institute of Theoretical Physics, Technische Universit\"{a}t Dresden, 01062 Dresden, Germany}
\affiliation{W\"{u}rzburg-Dresden Cluster of Excellence ct.qmat, Germany}

\author{Bj\"orn Trauzettel}
\affiliation{Institute of Theoretical Physics and Astrophysics, University of W\"{u}rzburg, Am Hubland, 97074 W\"{u}rzburg, Germany}
\affiliation{W\"{u}rzburg-Dresden Cluster of Excellence ct.qmat, Germany}

%\date{\today}

\begin{abstract}
We reveal a topological burning glass effect, where the local response of a system exhibits a topological quantization that is enhanced by an integer due to its environmental coupling. As a paradigmatic platform for this intriguing phenomenon, we study a central spin that is quasiperiodically driven by two incommensurate frequencies, and statically coupled to $N-1$ surrounding spins. In the strong coupling regime, the adiabatic dynamics of the total system is readily understood to imprint on the central spin an $N$-fold enhanced topological frequency conversion between the two driving frequencies. We argue that the topological burning glass effect is induced by the non-unitary dynamics of the central spin, which locally involves the collective motion of the surrounding spins. Our results are derived in the framework of adiabatic perturbation theory and fully corroborated by exact numerical simulations.
%\\[-13pt]\begin{center}{\it version~\input{|"git describe --tags"}}\\[-13pt]\end{center}
%\\[-13pt]\begin{center}{\it version~\input{version}}\\[-13pt]\end{center}
\end{abstract}

\pacs{}

%\keywords{}

\maketitle

{\it Introduction.}---Geometry and topology have been identified to be at the heart of some
of the most fascinating phenomena in physics~\cite{Thouless_1998,Xiao2010,Wen_2017,Cayssol_2021}. A paradigmatic example is
provided by the quantum Hall effect~\cite{Klitzing_1980,Prange_1990}, where an impressively precise quantization
of the transverse conductance in two-dimensional (2D) systems has been explained
with theory in terms of a topological invariant characterizing the underlying electronic state~\cite{Laughlin_1981,Thouless_1982,Simon_1983,Haldane1988}.

Generally speaking, linear response properties such as the aforementioned Hall conductance
are governed by the celebrated Kubo formula~\cite{Kubo_1957,Mahan_2000}. When interpreted as a dynamical expression,
linear response ostensibly refers to the first perturbative correction of the state due to
an external field. However, as an equilibrium property, it can eventually be understood
in terms of the unperturbed state alone. Returning to the integer quantum Hall effect, the topologically
quantized response is thus characterized by the first Chern number~\cite{Chern_1946,Thouless_1982,Qi_2008,Hasan2010,Qi_2011,Bernevig2013}, an invariant defined in terms of the adiabatic Berry curvature~\cite{Berry1984} of the occupied Bloch bands exclusively referring to the ground state of the system.

\begin{figure}
	\includegraphics[width=\columnwidth]{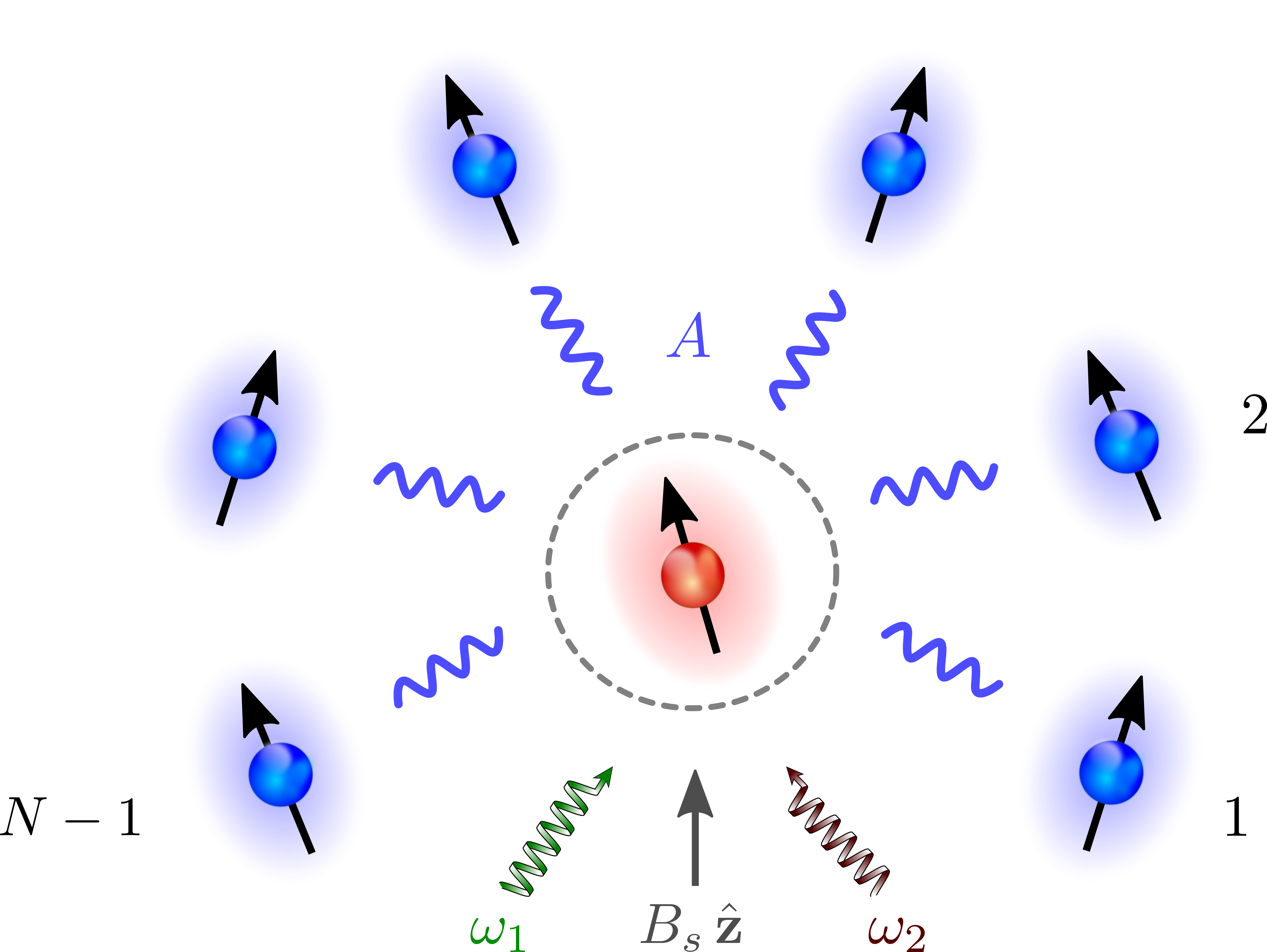}
	\caption{Quasiperiodically driven central spin model (CSM) as a topological burning glass. A central spin (red sphere) couples in star-geometry to $N-1$ surrounding spins (blue spheres). The isotropic interaction is parametrized by the homogeneous coupling constant $A$. The central spin is subjected to a static magnetic field with amplitude $B_s$ and two circularly polarized drives with incommensurate frequencies $\omega_1$ and $\omega_2$, so as to induce a topological frequency conversion.}
	\label{fig:Setup}
\end{figure}

In this work, we reveal how this correspondence between adiabatically defined topological invariants
and quantized response properties is fundamentally modified in open quantum systems. To this end, we extend the Floquet counterpart of an
integer quantum Hall system~\cite{Martin_PRX17}, i.e. a spin driven with two incommensurate frequencies, by statically
coupling it to a set of surrounding spins in the framework of a central spin model (CSM)~\cite{Gaudin_1976,Dukelsky_2004,Claeys_2019} (see Fig.~\ref{fig:Setup}). In this system, the topologically quantized response is given by the energy transfer between the two driving modes,
which represents a local observable of the driven central spin. Yet, the adiabatically defined Chern number
of the central spin fails to predict the linear response signal, which can instead only be topologically understood
from the adiabatic state of the total system. In this sense, the topological properties of the environmental spins are focused to the local response of the central spin, motivating the terminology of a topological burning glass. We argue that the topological burning glass effect generically occurs in any static extension of the CSM under the following three conditions: $(i)$ the initial state of the interacting spin system is energetically separated from the other bands, $(ii)$ the dynamics is well described within first-order adiabatic perturbation theory~\cite{Rigolin_2008,Weinberg_2017}, $(iii)$ the time-quasiperiodic fields only act on the central spin. In our concrete topological burning glass scenario, the quantized energy transfer is found to be $N$-fold enhanced by the collective motion of $N$ spins, while the adiabatic Chern number
of the driven spin remains equal to one. Our analytical results are corroborated by numerically exact simulations, which also allow us to systematically study the breakdown of the adiabatic dynamics in our model. This nonadiabatic breakdown occurs when the driving frequencies approach the size of the energy gap above the many-body ground state that generically scales as $1/N$ in our model.

{\it Model and adiabatic perturbation theory.}---We consider the dynamics of a central spin that is subjected to a time-quasiperiodic magnetic field $\mathbf{B}(\vec{\varphi}_t)$, and interacts in a static fashion with its environment. A minimal framework for this scenario is provided by the following driven CSM (see Fig.~\ref{fig:Setup} for an illustration):
\begin{equation}
\hat{H}(\vec{\varphi}_t)=g^\ast\,\mu_B\,\mathbf{B}(\vec{\varphi}_t)\cdot\hat{\mathbf{S}}_0-A\,\hat{\mathbf{S}}_0\cdot\hat{\mathbf{J}},
\label{eq:CentralSpinModel}
\end{equation}
where $\hat{\mathbf{S}}_0=\frac{1}{2}\,\hat{\boldsymbol{\sigma}}_0$ represents the central spin-1/2 and $\hat{\mathbf{J}}=\sum_{k=1}^{N-1}\,\hat{\mathbf{S}}_k=\sum_{k=1}^{N-1}\,\frac{\hat{\boldsymbol{\sigma}}_k}{2}$ the surrounding spins. The environment is composed of a number of $N-1$ spin-1/2, and is assumed to interact homogeneously with the central spin with coupling strength $A$. We have set $\hbar=1$ and introduced the vector of Pauli matrices $\hat{\boldsymbol{\sigma}}_i=(\hat{\sigma}_i^{x},\hat{\sigma}_i^{y},\hat{\sigma}_i^{z})$ acting on the individual subspaces of the constituents. The Zeeman term generated by the magnetic field $\mathbf{B}(\vec{\varphi}_t)$ is proportional to Bohr's magneton $\mu_B$ and the effective $g^\ast-$factor of the central spin. This model effectively applies for instance to driven electron spins trapped in lateral quantum dots in SiGe or nitrogen-vacancy (NV) centers in diamond~\footnote{In lateral quantum dots or NV centers, the Heisenberg interaction to the surrounding spins is not necessarily isotropic, as assumed for simplicity in Eq.~\eqref{eq:CentralSpinModel}. However, as our main statements about the topological burning glass effect reflect fundamental concepts that do not depend on the details of the model, our results generally apply to anisotropic interactions as well.}. In these systems, the (central) electron spin couples in a star-like fashion to the surrounding nuclear spins of the host material, with a hyperfine interaction that is typically three orders of magnitude larger than the dipole-dipole coupling between the nuclear spins themselves~\cite{Hanson_2007,Doherty_2013,Chatterjee_2021}. 

The time-dependence of the external field $\mathbf{B}(\vec{\varphi}_t)=B_c\,\mathbf{d}(\vec{\varphi}_t)$ is chosen as~\cite{Haldane1988,Qi2006,Martin_PRX17}
\begin{equation*}
\mathbf{d}(\vec{\varphi}_t)=\begin{pmatrix}
\sin(\varphi_{1t})\\
\sin(\varphi_{2t})\\
M-\cos(\varphi_{1t})-\cos(\varphi_{2t})
\end{pmatrix}.
\end{equation*}
It consists of a static magnetic field with amplitude $B_s=B_c\,M$ in $z$ direction, and two circularly polarized drives, with amplitudes $B_c$ and time-dependent phases $\vec{\varphi}_t=(\varphi_{1t},\varphi_{2t})=\vec{\omega}\,t+\vec{\phi}$. The offset phases and incommensurate frequencies are parametrized by $\vec{\phi}=(\phi_1,\phi_2)$ and $\vec{\omega}=(\omega_1,\omega_2)$.

In order to cope with the influence of the surrounding spins, we take advantage of the homogeneity of the interaction based on the star-like symmetry of Eq.~\eqref{eq:CentralSpinModel}, that entails $[\hat{H}(\vec{\varphi}_t),\hat{\mathbf{J}}^2]=0$. The Hamiltonian $\hat{H}(\vec{\varphi}_t)$ may thus be transformed into a block diagonal form, with each block characterized by a constant of motion
\begin{equation*}
j=\begin{cases}
\frac{N-1}{2},\,\frac{N-1}{2}-1,\,\dots,\,\frac{1}{2} & ;\,\text{$N$ even}\\[0.2cm]
\frac{N-1}{2},\,\frac{N-1}{2}-1,\,\dots,\,0 &  ;\,\text{$N$ odd}\\
\end{cases}
\end{equation*}
corresponding to the total spin $\hat{\mathbf{J}}^2$ of the surrounding spins with $N\geq2$. As we want to focus on the dynamics originating from the ground state, we restrict ourselves to the block $j=\frac{N-1}{2}$ (see Appendix~\ref{sec:InstantaneousSpectrum}). The associated block matrix size is $2N\times2N$, and the instantaneous spectrum is obtained by solving
\begin{equation}
\hat{H}(\vec{\varphi}_t)\,\ket{\Phi_\alpha(\vec{\varphi}_t)}=E_\alpha(\vec{\varphi}_t)\,\ket{\Phi_\alpha(\vec{\varphi}_t)}.
\label{eq:InstantaneousSpectrum}
\end{equation}
The energies $E_\alpha(\vec{\varphi}_t)$ are ordered from low to high values by ascending indices $\alpha=0,\,1,\,\dots,\,2N-1$, where $\alpha=0$ denotes the ground state. For an incommensurate pair of frequencies $\frac{\omega_1}{\omega_2}\notin\mathbb{Q}$, the phases $\varphi_{1t}$ and $\varphi_{2t}$ sample entirely the surface of a two-dimensional torus, analogously to a synthetic 2D Brillouin zone (BZ). The energy levels $E_\alpha(\vec{\varphi}_t)$ then resemble a band structure in parameter space, with the phases $\vec{\varphi}_t$ taking the role of Bloch quasimomenta. Assuming that the interacting spin system is gapped (see Appendix~\ref{sec:InstantaneousSpectrum}) and initially prepared in an instantaneous eigenstate $\ket{\Phi_\beta(\vec{\varphi}_{t_0})}$ of Eq.~\eqref{eq:InstantaneousSpectrum}, the dynamics can be expanded to first order in $\vec{\omega}$ using adiabatic perturbation theory~\cite{Rigolin_2008,Weinberg_2017}:
\begin{equation}
\ket{\Psi_\beta(t)}=\mathrm{e}^{\mathrm{i}\gamma_\beta(t)}\Big[\ket{\Phi_\beta(\vec{\varphi}_t)}+\mathrm{i}\sum_{\alpha\neq\beta}\frac{\mathcal{M}_{\alpha\beta}(\vec{\varphi}_t)}{\Delta_{\alpha\beta}(\vec{\varphi}_t)}\ket{\Phi_\alpha(\vec{\varphi}_t)}\Big].
\label{eq:AdiabaticPerturbationTheory}
\end{equation}
At zeroth order, the quantum state $\ket{\Psi_\beta(t)}$ is restricted to the synthetic energy band $E_\beta(\vec{\varphi}_t)$. First-order corrections, however, yield virtual transitions to the excited states of the instantaneous spectrum, weighted by the terms $\mathcal{M}_{\alpha \beta}(\vec{\varphi}_t)=\vec{\omega}\,\braket{\Phi_\alpha(\vec{\varphi}_t)|\nabla_{\vec{\varphi}}\,\Phi_\beta(\vec{\varphi}_t)}$ and the energy gaps
$\Delta_{\alpha\beta}(\vec{\varphi}_t)=E_\alpha(\vec{\varphi}_t)-E_\beta(\vec{\varphi}_t)$. We have also introduced the overall phase factor $\gamma_\beta(t)=-\int_{t_0}^{t}\mathrm{d}t^\prime\,[E_\beta(\vec{\varphi}_{t^\prime})-\mathrm{i}\,\mathcal{M}_{\beta\beta}(\vec{\varphi}_{t^\prime})]$.

{\it Topological frequency conversion.}---A feature of quasiperiodically driven systems is energy pumping, a process in which photons of different frequencies are exchanged between the external drives~\cite{Martin_PRX17,Kolodrubetz2018,Peng2018,Crowley2019,Nathan2019,Koerber_2020,Crowley_2020,Boyers_2020,Chen_2020,Nathan_2020,Long_2021,Nathan_2021,Psaroudaki_2021,Qi_2021,Schwennicke_2022}. The total energy transfer rate is determined by the equations of motion: $\frac{\mathrm{d}}{\mathrm{d}t}\braket{\hat{H}}=\braket{\partial_t\,\hat{H}}=\vec{\omega}\,\braket{\hat{\mathbf{I}}}$, with $\hat{\mathbf{I}}=(\hat{\text{I}}_1,\hat{\text{I}}_2)=\nabla_{\vec{\varphi}}\,\hat{H}$. Each term $\omega_k\braket{\hat{\text{I}}_k}$ can be interpreted as the pumping rate provided by the individual drive, where $\hat{\text{I}}_k=\partial_{\varphi_k}\hat{H}$ resembles a current operator in $k$ direction of the parameter space. Using the perturbed quantum state $\ket{\Psi_\beta(t)}$ from Eq.~\eqref{eq:AdiabaticPerturbationTheory}, the expectation value $\braket{\hat{\text{I}}_k}$ can be expanded to first order in $\vec{\omega}$~\cite{Gritsev_2012,Weinberg_2017}:
\begin{equation}
\braket{\hat{\text{I}}_k}_\beta=\braket{\Psi_\beta|\hat{\text{I}}_k|\Psi_\beta}=\frac{\partial E_\beta}{\partial\varphi_k}+\sum_{l=1}^{2}\,\omega_l\,\Omega^{(\beta)}_{kl}.
\label{eq:GeometricalConcepts}
\end{equation}
This result emphasizes that virtual couplings between bands are essential for the generation of geometrical and topological effects, as the adiabatic limit only produces Bloch oscillations $\partial_{\varphi_k}E_\beta$~\cite{Bloch_1929}. In fact, the virtual inter-band excitations of Eq.~\eqref{eq:AdiabaticPerturbationTheory} can be readily shown to be identical to the Berry curvature $\Omega^{(\beta)}_{kl}(\vec{\varphi}_t)=2\,\text{Im}[\braket{\partial_{\varphi_k}\,\Phi_\beta(\vec{\varphi}_t)|\partial_{\varphi_l}\,\Phi_\beta(\vec{\varphi}_t)}]$ of the synthetic energy band $E_\beta(\vec{\varphi}_t)$ to which the adiabatic dynamics is confined. This phenomenon demonstrates that, in an isolated quantum system, the Berry curvature $\Omega^{(\beta)}_{kl}$ arises as the adiabatic first-order response of a physical observable ($\hat{\text{I}}_k=\partial_{\varphi_k}\hat{H}$) to the rate of change of the external parameter (e.g. $\dot{\vec{\varphi}}_t=\vec{\omega}$)~\cite{Gritsev_2012,Weinberg_2017}.

As the two frequencies $\frac{\omega_1}{\omega_2}\notin\mathbb{Q}$ are incommensurate, the entire synthetic 2D BZ is sampled during the time-evolution of the quantum state $\ket{\Psi_\beta(t)}$. Averaging the pumping rates of Eq.~\eqref{eq:GeometricalConcepts} over a long period of time, this translates into an integration over the closed manifold of the two-dimensional torus. The Bloch oscillations do not contribute to this integration, while the integrated Berry curvature produces a topological frequency conversion between the dynamical drives that is proportional to the first Chern number $C^{(\beta)}=\frac{1}{2\pi}\iint_{0}^{2\pi}\mathrm{d}^2\vec{\varphi}\,\Omega^{(\beta)}_{12}(\vec{\varphi})$. The time-averaged pumping rate yields~\cite{Martin_PRX17}
\begin{equation}
P_\beta^{12}=-P_\beta^{21}=\frac{C^{(\beta)}}{2\,\pi}\,\omega_1\,\omega_2.
\label{eq:PumpingRate}
\end{equation}
The topologically quantized response of the total quantum system of Eq.~\eqref{eq:CentralSpinModel} is thus characterized by its adiabatically defined topological invariant $C^{(\beta)}$. We contrast this behavior with its counterpart in an open quantum system in the next section.

{\it Burning glass effect.}---We take an open quantum system perspective, in which a physical observable of interest operates locally on the (small) quantum system that is coupled to a (larger) environment. Specifically, in Eq.~\eqref{eq:CentralSpinModel}, only the central spin is exposed to the quasiperiodic field $\mathbf{B}(\vec{\varphi}_t)$, resulting in a current operator $\hat{\text{I}}_k=\lambda\,\frac{\partial\mathbf{d}(\vec{\varphi}_t)}{\partial \varphi_k}\cdot\hat{\mathbf{S}}_0$ that acts exclusively on the central spin-1/2. Accordingly, the expectation value
\begin{equation}
\braket{\hat{\text{I}}_k}_\beta=\Tr[\hat{\rho}_\beta^\text{dy}\,\hat{\text{I}}_k]=\Tr_0[\hat{\rho}^\text{dy}_{0,\beta}\,\hat{\text{I}}_k]
\label{eq:ExpectationValue}
\end{equation}
can be fully determined by the non-unitary dynamics of the central spin, expressed by the reduced density matrix $\hat{\rho}^\text{dy}_{0,\beta}(t)=\Tr_{\hat{\mathbf{J}}}[\hat{\rho}_\beta^\text{dy}(t)]$. Here, we have introduced the total density matrix $\hat{\rho}_\beta^\text{dy}(t)=\ketbra{\Psi_\beta(t)}{\Psi_\beta(t)}$, where $\ket{\Psi_\beta(t)}$ represents the perturbed quantum state of Eq.~\eqref{eq:AdiabaticPerturbationTheory}~\footnote{Note that Eq.~\eqref{eq:ExpectationValue} also holds for the exact dynamics of the system, involving the higher orders in the perturbative expansion of Eq.~\eqref{eq:AdiabaticPerturbationTheory}.}. The reduced density matrix is calculated by tracing out the environment $\hat{\mathbf{J}}$ (denoted as $\Tr_{\hat{\mathbf{J}}}$), while $\Tr$ ($\Tr_0$) denotes the trace operating on the total system (the central spin-1/2). We have further introduced the energy scale
\begin{equation*}
\lambda=g^\ast\,\mu_B\,B_c,
\end{equation*} 
assuming $\lambda>0$ for simplicity.

Eq.~\eqref{eq:ExpectationValue} illustrates that the frequency conversion is entirely carried by the central spin. The topological quantization of the local response, however, is determined by the geometrical and topological properties of the total system, namely the Berry curvature $\Omega^{(\beta)}_{kl}$ and the Chern number $C^{(\beta)}$ of the synthetic energy band $E_\beta(\vec{\varphi}_t)$ to which the adiabatic dynamics is confined. Thus, the non-unitary dynamics of the central spin effectively inherits the topological nature of the total system. Employing the perturbed quantum state $\ket{\Psi_\beta(t)}$ of Eq.~\eqref{eq:AdiabaticPerturbationTheory}, this phenomenon can be further analyzed by a first-order expansion in $\vec{\omega}$ of the reduced density matrix 
\begin{equation}
\hat{\rho}^\text{dy}_{0,\beta}(t)=\hat{\rho}^\text{ad}_{0,\beta}(t)+\sum_{\alpha\neq\beta}\frac{\hat{T}_{\alpha\beta}(\vec{\varphi}_t)}{\Delta_{\alpha\beta}(\vec{\varphi}_t)}.
\label{eq:ReducedDensityMatrix}
\end{equation}
Here, we introduce the adiabatic limit of the reduced density matrix $\hat{\rho}^\text{ad}_{0,\beta}=\Tr_{\hat{\mathbf{J}}}[\ketbra{\Phi_\beta}{\Phi_\beta}]$, which corresponds to the zeroth order of Eq.~\eqref{eq:AdiabaticPerturbationTheory}, and results in Bloch oscillations $\partial_{\varphi_k}E_\beta$ of Eq.~\eqref{eq:GeometricalConcepts}. The local response, however, arises from the operators $\hat{T}_{\alpha\beta}=\mathrm{i}\,\mathcal{M}_{\alpha \beta}\,\Tr_{\hat{\mathbf{J}}}[\ketbra{\Phi_\alpha}{\Phi_\beta}]+\text{h.c.}$ accounting for the virtual transitions to the excited states in the instantaneous spectrum of the total system. In fact, the correlations of the total system are manifested in the matrix elements of $\hat{T}_{\alpha\beta}$, thus imposing a topological quantization that is not captured by basic geometrical or topological aspects of the reduced adiabatic density matrix $\hat{\rho}^\text{ad}_{0,\beta}$. In this sense, the adiabatic topological properties of the total system are focused to the local response of the central spin, which motivates the terminology of a topological burning glass. This mechanism is quite generic, since Eqs.~\eqref{eq:InstantaneousSpectrum} to \eqref{eq:ReducedDensityMatrix} reflect fundamental concepts that do not depend on the details of the model. Instead, it is only assumed that the adiabatic dynamics starts from a gapped synthetic energy band of the total system, and that the physical observable of interest operates locally on the (small) quantum system. The robustness of our results to more generic couplings is further corroborated in the Appendix~\ref{sec:StabilityAnalysis}.

To further analyze the topological burning glass effect, we contrast the dynamics of a decoupled central spin with the collective dynamics in the strong coupling regime. We start with the noninteracting case, where Eq.~\eqref{eq:CentralSpinModel} transforms into a single-spin Hamiltonian resembling the momentum-space representation of a Chern insulator~\cite{Haldane1988,Qi2006} with mass parameter $M$. Thus, each of the two single-spin energy bands can be characterized by a Chern number~\cite{Chern_1946,Thouless_1982,Qi_2008,Hasan2010,Qi_2011,Bernevig2013}: $\nu_{gr}=-\nu_{ex}=\pm1$ (nontrivial) for $|M|<2$, $M\neq0$ or $\nu_{gr}=\nu_{ex}=0$ (trivial) for $|M|>2$, where $\nu_{gr}$ ($\nu_{ex}$) corresponds to the single-spin ground (excited) energy band. Starting from the single-spin ground state, the adiabatic dynamics produces a topological frequency conversion that is proportional to $	\nu_{gr}=-\frac{1}{4\pi}\iint_{0}^{2\pi}\mathrm{d}^2\vec{\varphi}\,\tilde{\mathbf{d}}(\vec{\varphi})\,\Big(\partial_{\varphi_1}\tilde{\mathbf{d}}(\vec{\varphi})\times\partial_{\varphi_2}\tilde{\mathbf{d}}(\vec{\varphi})\Big) $ with $\tilde{\mathbf{d}}(\vec{\varphi})=\frac{\mathbf{d}(\vec{\varphi})}{|\mathbf{d}(\vec{\varphi})|}$. In the interacting case, the interaction effectively extends the magnetic coupling to the surrounding spins, forcing them to rotate along the direction of the external field $\mathbf{B}(\vec{\varphi}_t)$ as well. In the strong coupling regime and for ferromagnetic coupling strength $A>0$~\footnote{Results for antiferromagnetic coupling strength $A<0$ are presented in the Appendix~\ref{sec:AntiferromagneticInteraction}}, this collective behavior imposes a topological frequency conversion that is proportional (cf.~Eq.~\eqref{eq:PumpingRate}) to the total Chern number
\begin{equation}
C^{(0)}=N\,\nu_{gr}
\label{eq:EnhancedConversion}
\end{equation}
of the ferromagnetic ground state. This behavior affords a simple interpretation: In the adiabatic limit, the surrounding spins point in the same direction as the central spin at every point in time, such that each spin contributes to the many-body wave function by the same single-spin Chern number $\nu_{gr}$. Importantly, the adiabatic ground state is a product state at all times, such that the reduced adiabatic density matrix $\hat{\rho}^\text{ad}_{0,0}=\frac{1}{2}\,(\mathbbm{1}-\tilde{\mathbf{d}}(\vec{\varphi})\cdot\hat{\boldsymbol{\sigma}})$ yields a Chern number $\nu_{gr}$ that is given by that of a single spin-1/2. The latter thus fails to capture the topological quantization and enhancement of the local response, revealing a burning glass effect in which the aforementioned collective motion of all spins is locally imprinted in the non-unitary dynamics of the central spin. In fact, as we describe in the Appendix~\ref{sec:DeflectionsSpinPolarization}, the operators $\hat{T}_{\alpha\beta}$ of Eq.~\eqref{eq:ReducedDensityMatrix} induce fluctuations around the adiabatic spin polarization of the central spin-1/2, including the topological information of the total system, and affecting the indirect measurement of the topological frequency conversion~\cite{Schroer_2014,Roushan2014,Malz_2021}.

\begin{figure}
	\includegraphics[width=\columnwidth]{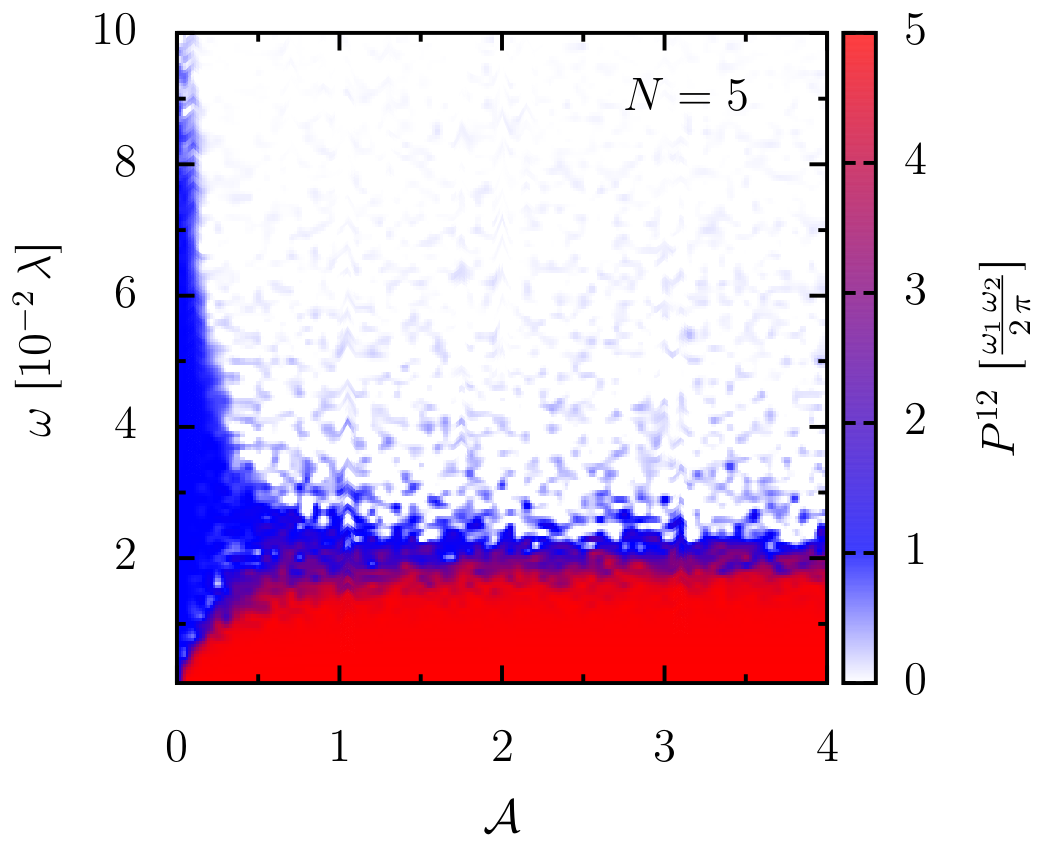}	
	\caption{Nonequilibrium phase diagram as a function of interaction strength $\mathcal{A}>0$ and frequencies $\omega_1=\omega$, $\omega_2=\gamma\,\omega$ for $N=5$. The mass parameter $M=1.2$ is chosen in the nontrivial topological regime ($\nu_{gr}=1$). Provided the adiabatic dynamics is confined to the ferromagnetic ground state, the time-averaged pumping rate $P^{12}$ is proportional to the Chern number $C^{(0)}=N\,\nu_{gr}$ (red regime). Nonadiabatic excitations can result in transitions to intermediate/featureless quantum phases with pumping rates $\bar{P}^{12}=\frac{\nu_{gr}}{2\,\pi}\,\omega_1\,\omega_2$ (blue regime)/$P^{12}=0$ (white regime).}
	\label{fig:PhaseDiagram}
\end{figure}

{\it Numerical simulations.}---We corroborate the formation of a topological burning glass by numerically solving the Schrödinger equation associated with $\hat{H}$ up to times $\lambda\,T=5\cdot10^5$~\footnote{For a time $\lambda\,T=5\cdot10^5$, the entire synthetic 2D Brillouin zone (BZ) is sufficiently sampled for our choice of parameters.}. If the initial state at $t_0=0$ corresponds to the ferromagnetic ground state, the time-averaged pumping rate $P^{12}$ can be extrapolated by the gradient~\footnote{We obtain the gradient by computing $P^{12}=\frac{2}{T}\,(\bar{W}_{1}^{T}-\bar{W}_{1}^{T/2})$ with the time-averaged energy transfers $\bar{W}_{1}^{T/2}=\frac{2}{T}\,\int_{0}^{T/2}\,\mathrm{d}t\,W_{1}(t)$ and $\bar{W}_{1}^{T}=\frac{2}{T}\,\int_{T/2}^{T}\,\mathrm{d}t\,W_{1}(t)$.} of the energy transfer $W_{1}(t)=\omega_{1}\,\int_{0}^{t}\,\mathrm{d}t^\prime\,\braket{\hat{\text{I}}_{1}(t^\prime)}$. In Fig.~\ref{fig:PhaseDiagram}, the pumping rate $P^{12}$ is shown as a function of ferromagnetic interaction strength $\mathcal{A}>0$~\cite{Note3} and frequency $\omega$ for a total number of spins $N=5$, where we have introduced the renormalized interaction strength $\mathcal{A}=A/\lambda$. The mass parameter $M=1.2$ is selected such that the system is in the nontrivial topological regime, yielding an adiabatic Chern number $\nu_{gr}=1$ of the driven spin. We have chosen the frequencies $\omega_1=\omega$ and $\omega_2=\gamma\,\omega$, where $\gamma=\frac{1}{2}(1+\sqrt{5})$ is the golden ratio. The offset phases are $\phi_1=\pi/10$, $\phi_2=0$.

A finite interaction strength drives the system into the burning glass regime with an $N$-fold enhanced topological frequency conversion determined by the Chern number of Eq.~\eqref{eq:EnhancedConversion} (red regime). Away from the zero-frequency limit, there exists parameter ranges at which the strong coupling regime of Eq.~\eqref{eq:PumpingRate} breaks down. The dynamics leads to nonadiabatic excitation processes between the instantaneous eigenstates of the spectrum, yielding a suppression of the dynamical response. Finally, the system enters an ergodic regime that leads to a featureless state with pumping rate $P^{12}=0$ (white regime). Interestingly, in the CSM, there exists an intermediate dynamical quantum phase in which the spectrum is only partially occupied. The topological frequency conversion then even extends to a nonadiabatic situation, with a pumping rate $\bar{P}^{12}=\frac{\nu_{gr}}{2\,\pi}\,\omega_1\,\omega_2$ similar to that of the single spin (blue regime). A detailed description of the nonadiabatic breakdown of the strong coupling regime is presented in the Appendix~\ref{sec:BreakdownStrongCouplingRegime}, showing that the parameter ranges of the nonequilibrium phase diagram mainly depend on the relations between the driving frequencies and the energy gaps of the band structure. More precisely, the phase boundaries in Fig.~\ref{fig:PhaseDiagram} are approximately covered by critical frequencies proportional to the energy gaps separating the spectrum into the aforementioned dynamical quantum phases. Note that these energy gaps scale with $1/N$, indicating that the topological burning glass effect represents a finite size effect in our model.

{\it Conclusion.}---We have discovered a topological burning glass effect, in which the local response of a quantum system exhibits a topological quantization that is enhanced by an integer due to its environmental coupling. The quantum system adopts the topological nature of the total system in its non-unitary dynamics, imposing a local response that can not be captured by its reduced adiabatic density matrix. We expect that this intriguing phenomenon exemplifies a more general principle of topological open quantum systems.

As a prototypical example of a topological burning glass, we have studied energy pumping in a quasiperiodically driven central spin model. In the strong coupling regime, the central spin experiences a magnification of the topological frequency conversion that is significantly enhanced with the number of surrounding spins. This amplification could be exploited to enable the direct experimental observation of a quantized energy current, providing a complementary approach to the recent suggestions made in the context of Weyl semimetals~\cite{Nathan_2022}.

\begin{acknowledgments}
This work was supported by the DFG [SPP1666, SFB1170 "ToCoTronics", SFB 1143 (project-id 247310070), and DFG Project No. 419241108], the W\"{u}rzburg-Dresden Cluster of Excellence ct.qmat (EXC2147, Project No. 390858490), and the Elitenetzwerk Bayern Graduate School on "Topological Insulators". The computational work presented in this manuscript was performed on the W\"{u}rzburg HPC cluster.
\end{acknowledgments}

\bibliographystyle{apsrev4-1}
%\bibliography{ThesisBiblio,DCSMNotes}

\onecolumngrid
\setcounter{secnumdepth}{2}
\appendix
\renewcommand{\thesection}{\Alph{section}}
\renewcommand{\thesubsection}{\Alph{section}\arabic{subsection}}
\renewcommand{\thefigure}{\Alph{section}\arabic{subsection}.\arabic{figure}}
\numberwithin{figure}{subsection}
\renewcommand{\theequation}{\Alph{section}\arabic{subsection}.\arabic{equation}}
\numberwithin{equation}{subsection}
\renewcommand{\thetable}{\Alph{section}\arabic{subsection}.\arabic{table}}
\numberwithin{table}{subsection}
\makeatletter
\renewcommand{\p@subsection}{}
\makeatother

\section*{Appendix}
In the main text, we have illustrated the topological burning glass effect by exploring energy pumping in the quasiperiodically driven central spin model (CSM). The main goal of this Appendix is to provide additional technical details about this prototypical example. This includes the derivation of the instantaneous spectrum of the interacting spin system (Appendix~\ref{sec:InstantaneousSpectrum}), the stability analysis for more generic extensions of the CSM (Appendix~\ref{sec:StabilityAnalysis}), the illustration of the fluctuations around the adiabatic spin polarization of the central spin-1/2 (Appendix~\ref{sec:DeflectionsSpinPolarization}), the detailed description of the nonadiabatic breakdown of the strong coupling regime with the associated scaling behavior (Appendix~\ref{sec:BreakdownStrongCouplingRegime}), and the results for antiferromagnetic coupling strength $A<0$ (Appendix~\ref{sec:AntiferromagneticInteraction}).    

\section{Instantaneous spectrum} \label{sec:InstantaneousSpectrum}
We analyze the block diagonal form of the total system and show that the overall ground state is contained in the block $j=\frac{N-1}{2}$ (Sec.~\ref{subsec:BlockDiagonalForm}). We restrict ourselves to this block and derive the topological band structure in the synthetic 2D Brillouin zone (BZ). This results in a gapped spectrum that is robust to variations in the interaction strength (Sec.~\ref{subsec:TopologicalBandStructure}).

\subsection{Block diagonal form} \label{subsec:BlockDiagonalForm}
Rewriting Eq.~\eqref{eq:CentralSpinModel} of the main text by means of the total spin $\hat{\mathbf{S}}_\text{tot}=\hat{\mathbf{S}}_0+\hat{\mathbf{J}}$, it becomes clear that the interaction is invariant under the transformation
\begin{equation}
\hat{H}^\prime(\vec{\varphi})=\Big(\bigotimes_{i=0}^{N-1}\,\hat{U}^\dagger(\vec{\varphi})\Big)\,\hat{H}(\vec{\varphi})\,\Big(\bigotimes_{i=0}^{N-1}\,\hat{U}(\vec{\varphi})\Big)=\lambda\,\Big[|\mathbf{d}(\vec{\varphi})|\,\hat{S}^{z^\prime}_0-\frac{\mathcal{A}}{2}\,\Big(\hat{\mathbf{S}}^2_\text{tot}-\hat{\mathbf{J}}^2-\frac{3}{4}\Big)\Big].
\label{eq:CSMRotated}
\end{equation} 
The $SU(2)$ matrices $\hat{U}(\vec{\varphi})$ rotate the $z$ quantization axes of the subspaces in the direction of the field vector $\mathbf{d}(\vec{\varphi})$: $\hat{U}^\dagger(\vec{\varphi})\,\mathbf{d}(\vec{\varphi})\cdot\hat{\boldsymbol{\sigma}}\,\hat{U}(\vec{\varphi})=|\mathbf{d}(\vec{\varphi})|\,\hat{\sigma}^{z^\prime}$. We have introduced the energy scale $\lambda=g^\ast\,\mu_B\,B_c$, the  renormalized interaction strength $\mathcal{A}=A/\lambda$, and the $z^\prime$ direction within the rotated basis. In what follows, we assume positive energy scales $\lambda>0$ for simplicity. By introducing raising and lowering operators $\hat{S}^{\prime\pm}_0=\hat{S}^{x^\prime}_0\pm\mathrm{i}\,\hat{S}^{y^\prime}_0$ and $\hat{J}^{\prime\pm}=\hat{J}^{x^\prime}\pm\mathrm{i}\,\hat{J}^{y^\prime}$ within the rotated basis, Eq.~\eqref{eq:CSMRotated} can be rewritten as
\begin{equation}
\hat{H}^\prime(\vec{\varphi})=\lambda\,\Big[|\mathbf{d}(\vec{\varphi})|\,\hat{S}^{z^\prime}_0-\frac{\mathcal{A}}{2}\,\Big(\hat{S}^{\prime+}_0\,\hat{J}^{\prime-}+\hat{S}^{\prime-}_0\,\hat{J}^{\prime+}+2\,\hat{S}^{z^\prime}_0\,\hat{J}^{z^\prime}\Big)\Big].
\label{eq:CSMSimplified}
\end{equation}
Eq.~\eqref{eq:CSMSimplified} immediately reveals the symmetries of the model: $[\hat{H}^\prime(\vec{\varphi}),\hat{\mathbf{J}}^2]=[\hat{H}(\vec{\varphi}),\hat{\mathbf{J}}^2]=0$ and $[\hat{H}^\prime(\vec{\varphi}),\hat{S}^{z^\prime}_\text{tot}]=0$. The associated quantum numbers are given by
\begin{equation*}
	j=\begin{cases}
		\frac{N-1}{2},\,\frac{N-1}{2}-1,\,\dots,\,\frac{1}{2} &  ;\,\text{$N$ even}\\[0.2cm]
		\frac{N-1}{2},\,\frac{N-1}{2}-1,\,\dots,\,0 & ;\,\text{$N$ odd}\\
	\end{cases}\quad \text{and} \quad m_s=-j-\frac{1}{2},\,-j+\frac{1}{2},\,\dots,\,j+\frac{1}{2}.
\end{equation*}
Due to the symmetries of the Hamiltonian $\hat{H}^\prime(\vec{\varphi})$, it seems convenient to represent Eq.~\eqref{eq:CSMSimplified} within the rotated basis of product states between the central spin-1/2 and the total spin $\hat{\mathbf{J}}^2$ of the surrounding spins: $\ket{m_{s_0},j,m_j,\eta}=\ket{m_{s_0}}\otimes\ket{j,m_j,\eta}$. We have introduced the eigenstates within the rotated basis  
\begin{equation}
\hat{S}^{z^\prime}_0\,\ket{m_{s_0}}=m_{s_0}\,\ket{m_{s_0}},\quad \hat{\mathbf{J}}^2\,\ket{j,m_j,\eta}=j\,(j+1)\,\ket{j,m_j,\eta},\quad \hat{J}^{z^\prime}\,\ket{j,m_j,\eta}=m_j\,\ket{j,m_j,\eta},
\label{eq:EigenstatesSpinOperators}
\end{equation}
with quantum numbers $m_{s_0}=\pm\frac{1}{2}$, $m_j=-j,\,-j+1,\,\dots,\,j$, and $\eta=1,\,2,\,\dots,\,\eta_{\text{max}}$. The quantum numbers $\eta$ distinguish orthogonal states $\braket{j,m_j,\eta|j,m_j,\tilde{\eta}}=\delta_{\eta,\tilde{\eta}}$, having identical quantum numbers $j$ and $m_j$. Consequently, $\eta_{\text{max}}$ equals the dimension of the degenerate subspace formed by states $\ket{j,m_j,\eta}$. The rotated basis states $\ket{m_{s_0},j,m_j,\eta}$ form an orthonormal set with 
\begin{equation}
\braket{m_{s_0},j,m_j,\eta|\tilde{m}_{s_0},\tilde{j},\tilde{m}_j,\tilde{\eta}}=\delta_{m_{s_0},\tilde{m}_{s_0}}\,\delta_{j,\tilde{j}}\,\delta_{m_j,\tilde{m}_j}\,\delta_{\eta,\tilde{\eta}}.
\label{eq:OrthonormalBasis}
\end{equation}
We would like to emphasize that the quantum numbers $j$ are constants of motion, while the quantum numbers $m_s$ are not. This follows from the fact that $\hat{S}^{z^\prime}_\text{tot}$ becomes explicitly time-dependent when rotated back into the original basis of Eq.~\eqref{eq:CentralSpinModel} of the main text. Dynamical excitations into states of different $m_s$ are then possible during the nonequilibrium evolution of the system. Concerning the dynamics, we can thus restrict ourselves to the basis states with fixed quantum numbers $j$. The Hamiltonian $\hat{H}^\prime(\vec{\varphi})$ is transformed into a block diagonal form, with each block characterized by the constants of motion $j$. The explicit representation of the blocks within the rotated basis $\braket{m_{s_0},j,m_j,\eta|\hat{H}^\prime(\vec{\varphi})|\tilde{m}_{s_0},j,\tilde{m}_j,\tilde{\eta}}$ can be derived using the properties of Eqs.~\eqref{eq:EigenstatesSpinOperators}, \eqref{eq:OrthonormalBasis} and 
\begin{equation*}
	\hat{S}^{\prime\pm}_0\,\ket{\pm\frac{1}{2}}=0,\quad \hat{S}^{\prime\pm}_0\,\ket{\mp\frac{1}{2}}=\ket{\pm\frac{1}{2}},\quad \hat{J}^{\prime\pm}\,\ket{j,m_j,\eta}=\sqrt{j\,(j+1)-m_j\,(m_j\pm 1)}\,\ket{j,m_j\pm 1,\eta}.
\end{equation*}
This yields the only nonvanishing matrix elements
\begin{equation}
\braket{m_{s_0},j,m_j,\eta|\hat{H}^\prime(\vec{\varphi})|m_{s_0},j,m_j,\eta}=\lambda\,m_{s_0}\,(|\mathbf{d}(\vec{\varphi})|-\mathcal{A}\,m_j)
\label{eq:MatrixElementsDiag}
\end{equation}
and
\begin{equation}
\braket{\mp\frac{1}{2},j,m_j\pm 1,\eta|\hat{H}^\prime(\vec{\varphi})|\pm\frac{1}{2},j,m_j,\eta}=-\frac{\lambda\,\mathcal{A}}{2}\,\sqrt{j\,(j+1)-m_j\,(m_j\pm 1)}.
\label{eq:MatrixElementsOff}
\end{equation}
By proper arrangement of the nonvanishing matrix elements of Eqs.~\eqref{eq:MatrixElementsDiag} and \eqref{eq:MatrixElementsOff}, the block $j$ can be further separated into one- and two-dimensional blocks characterized by quantum numbers $m_s$ (and $\eta$). It should be noted that the Hamiltonian $\hat{H}^\prime(\vec{\varphi})$ is degenerate in the quantum numbers $\eta$. The one-dimensional blocks correspond to eigenstates $\ket{E,j,m_s,\eta}$ with minimal and maximal values $m_s=\pm\frac{1}{2}\pm j$: 
\begin{equation}
\ket{E^{(1)}_\pm,j,\pm\frac{1}{2}\pm j,\eta}=\ket{\pm\frac{1}{2},j,\pm j,\eta},\quad E^{(1)}_\pm(\vec{\varphi},j)=\pm\frac{\lambda}{2}\Big[|\mathbf{d}(\vec{\varphi})|\mp \mathcal{A}\,j\Big].
\label{eq:Solutions1D}	
\end{equation}
We have introduced quantum numbers $E$ corresponding to the energy eigenvalues of the associated eigenstates $\ket{E,j,m_s,\eta}$. The remaining quantum numbers $m_s=-j+\frac{1}{2},\,-j+\frac{3}{2},\,\dots,\,j-\frac{1}{2}$ define two-dimensional blocks within the rotated basis $\{\ket{\frac{1}{2},j,m_s-\frac{1}{2},\eta},\ket{-\frac{1}{2},j,m_s+\frac{1}{2},\eta}\}$, resulting in equations
\begin{equation}
\lambda\,\Big(\frac{\mathbf{l}\,\hat{\boldsymbol{\sigma}}^\prime}{2}+\frac{\mathcal{A}}{4}\Big)\,\ket{E^{(2)}_\pm,j,m_s,\eta}=E^{(2)}_\pm(\vec{\varphi},j,m_s)\,\ket{E^{(2)}_\pm,j,m_s,\eta}
\label{eq:2DBlock}
\end{equation}
with
\begin{equation*}
	\mathbf{l}(\vec{\varphi},j,m_s)=\begin{pmatrix}
		-\mathcal{A}\,\sqrt{j\,(j+1)-(m_s^2-\frac{1}{4})}\\
		0\\
		|\mathbf{d}(\vec{\varphi})|-\mathcal{A}\,m_s
	\end{pmatrix}.
\end{equation*}
The solutions of Eq.~\eqref{eq:2DBlock} are given by
\begin{equation}
\ket{E^{(2)}_\pm,j,m_s,\eta}=\begin{cases}
\ket{\pm\frac{1}{2},j,m_s\mp\frac{1}{2},\eta} &  ;\,\mathcal{A}=0\\[0.2cm]
\frac{|\mathbf{l}|\pm\mathbf{l}\,\hat{\boldsymbol{\sigma}}^\prime}{\sqrt{2\,|\mathbf{l}|\,(|\mathbf{l}|\pm l_{z})}}\,\begin{pmatrix}
1\\
0
\end{pmatrix} &  ;\,\mathcal{A}\neq0\\
\end{cases},\quad
E^{(2)}_\pm(\vec{\varphi},j,m_s)=\pm\frac{\lambda}{2}\,\Big[|\mathbf{l}(\vec{\varphi},j,m_s)|\pm\frac{\mathcal{A}}{2}\Big].
\label{eq:Solutions2D}
\end{equation}

Assuming that $|\mathbf{d}(\vec{\varphi})|\neq0$, $\mathcal{A}\neq0$, and $\lambda>0$, we examine the overall ground state of Eqs.~\eqref{eq:Solutions1D} and \eqref{eq:Solutions2D}. By formally performing the derivative
\begin{equation*}
	\frac{\partial E^{(2)}_-(\vec{\varphi},j,m_s)}{\partial m_s}=\frac{\lambda\,|\mathbf{d}(\vec{\varphi})|\,\mathcal{A}}{2\,|\mathbf{l}(\vec{\varphi},j,m_s)|}
\end{equation*}
at a given point of the synthetic 2D BZ, we can conclude that for ferromagnetic/antiferromagnetic interaction strength $\mathcal{A}>0$/$\mathcal{A}<0$ the lowest energy level $E^{(2)}_-$ within each block $j$ is given by the minimal/maximal quantum number $m_s=-j+\frac{1}{2}$/$m_s=j-\frac{1}{2}$. Moreover, it can be shown that
\begin{equation*}
	\lim\limits_{m_s\mapsto-\frac{1}{2}-j}\,E^{(2)}_-(\vec{\varphi},j,m_s)=\frac{\lambda}{4}\,\bigg(\mathcal{A}-\Big\vert2\,|\mathbf{d}(\vec{\varphi})|+(2\,j+1)\,\mathcal{A}\Big\vert\bigg)\begin{cases}
		=E^{(1)}_-(\vec{\varphi},j) &  ;\,\mathcal{A}>0\\[0.2cm]
		\leq E^{(1)}_-(\vec{\varphi},j) & ;\,\mathcal{A}<0\\
	\end{cases},
\end{equation*}
which implies that the local ground state energy level within each block $j$ is given by $E^{(1)}_-(\vec{\varphi},j)$/$E^{(2)}_-(\vec{\varphi},j,j-\frac{1}{2})$ for ferromagnetic/antiferromagnetic interaction strength $\mathcal{A}>0$/$\mathcal{A}<0$. By additionally performing the derivatives 
\begin{equation*}
	\frac{\partial E^{(1)}_-(\vec{\varphi},j)}{\partial j}=-\frac{\lambda\,\mathcal{A}}{2},\quad \frac{\partial E^{(2)}_-(\vec{\varphi},j,j-\frac{1}{2})}{\partial j}=\frac{\lambda\,\mathcal{A}\,[2\,|\mathbf{d}(\vec{\varphi})|-(2\,j+1)\,\mathcal{A}]}{4\,|\mathbf{l}(\vec{\varphi},j,m_s)|},
\end{equation*}
it becomes evident that the block $j=\frac{N-1}{2}$ contains the overall ground state for interaction strengths $\mathcal{A}\neq0$. The associated instantaneous spectrum is not degenerate ($\eta_{\text{max}}=1$).

\subsection{Topological band structure} \label{subsec:TopologicalBandStructure}
Since the Hamiltonian $\hat{H}^\prime(\vec{\varphi})$ of Eq.~\eqref{eq:CSMRotated} commutes with $\hat{S}^{z^\prime}_\text{tot}$, the block $j=\frac{N-1}{2}$ can be further separated into one- and two-dimensional blocks characterized by quantum numbers $m_s=-\frac{N}{2},\,-\frac{N}{2}+1,\,\dots,\,\frac{N}{2}$. The one-dimensional blocks correspond to the minimal and maximal values $m_s=\pm N/2$, resulting in energy bands (see Eq.~\eqref{eq:Solutions1D})
\begin{equation}
E^{(1)}_{\pm}(\vec{\varphi})=\pm\frac{\lambda}{4}\,\Big[2\,|\mathbf{d}(\vec{\varphi})|\mp \mathcal{A}\,(N-1)\Big].
\label{eq:Energy1}
\end{equation}
The remaining quantum numbers $m_s$ define two-dimensional blocks with energy bands (see Eq.~\eqref{eq:Solutions2D})
\begin{equation}
E^{(2)}_\pm(\vec{\varphi},m_s)=\pm\frac{\lambda}{4}\,\Big[\sqrt{\mathcal{A}^2\,N^2+4\,|\mathbf{d}(\vec{\varphi})|\,(|\mathbf{d}(\vec{\varphi})|-2\,\mathcal{A}\,m_s)}\pm\mathcal{A}\Big].
\label{eq:Energy2}
\end{equation}
The spectrum of Eqs.~\eqref{eq:Energy1} and \eqref{eq:Energy2} is gapped throughout the entire synthetic 2D BZ for $|\mathbf{d}(\vec{\varphi})|\neq0$ and $\mathcal{A}\neq0$. The dependence on the phases $\vec{\varphi}$ can then be addressed by introducing the ($\vec{\varphi}$-dependent) interaction strength $\mathcal{A}_{\vec{\varphi}}=\mathcal{A}/|\mathbf{d}(\vec{\varphi})|$ and the dimensionless energy eigenvalues
\begin{equation}
\mathcal{E}^{(1)}_\pm(\mathcal{A}_{\vec{\varphi}})=\frac{E^{(1)}_\pm(\vec{\varphi})}{\lambda\,|\mathbf{d}(\vec{\varphi})|}, \quad \mathcal{E}^{(2)}_{\pm,n}(\mathcal{A}_{\vec{\varphi}})=\frac{E^{(2)}_\pm(\vec{\varphi},-\frac{N}{2}+n)}{\lambda\,|\mathbf{d}(\vec{\varphi})|}. 
\label{eq:Energies}
\end{equation}
The eigenvalues $\mathcal{E}^{(1)}_\pm$ and $\mathcal{E}^{(2)}_{\pm,n}$ depend only on the interaction strength $\mathcal{A}_{\vec{\varphi}}$, but not on the explicit phases $\vec{\varphi}$. Instead of the quantum numbers $m_s$, we have introduced the number $n=1,\,2,\,\dots,\,N-1$ of flipped spins as compared to the ferromagnetic state $m_s=-\frac{N}{2}$ for illustration. 

\begin{figure}
	\includegraphics[width=0.6\columnwidth]{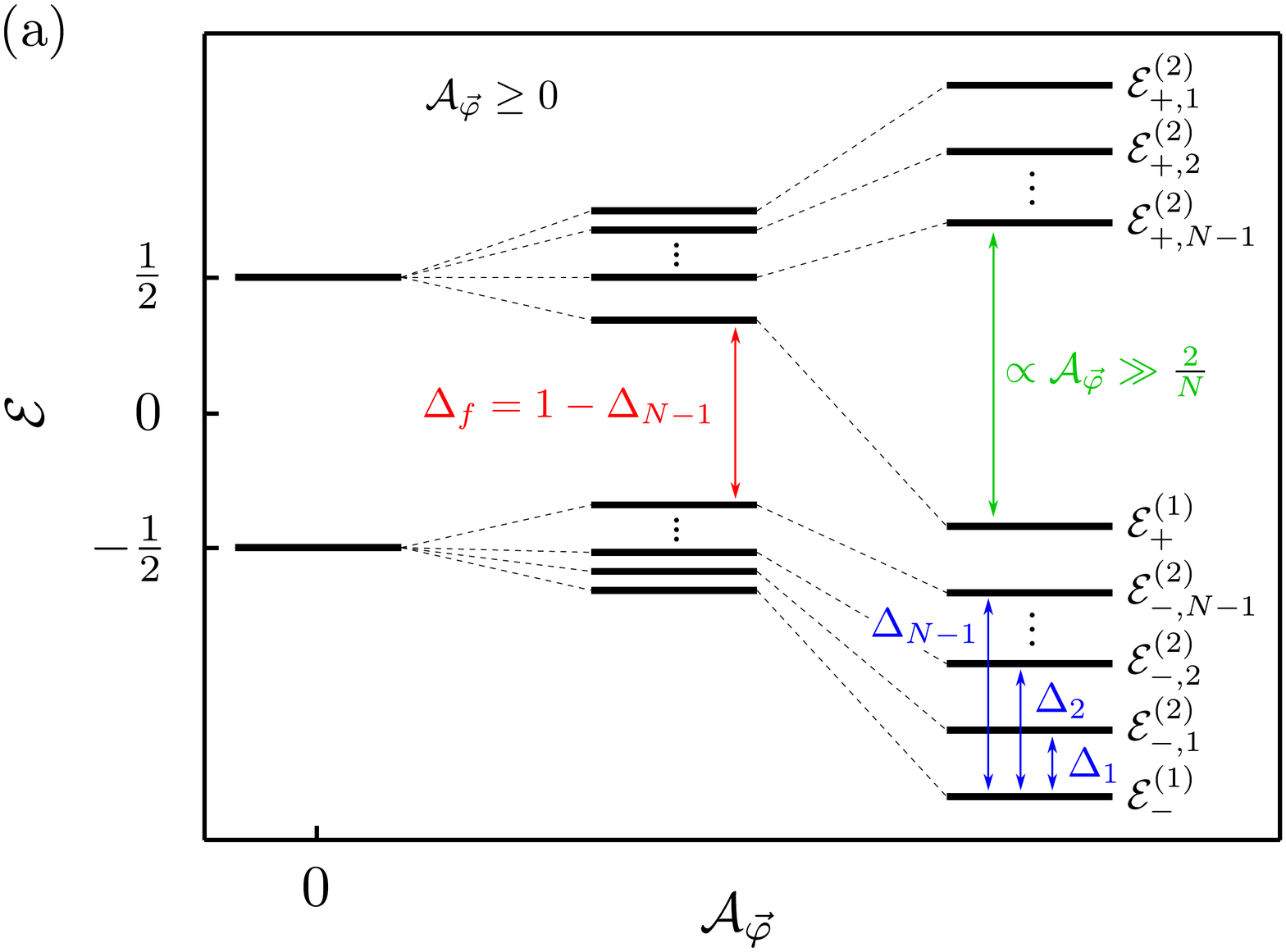}
	\includegraphics[width=0.6\columnwidth]{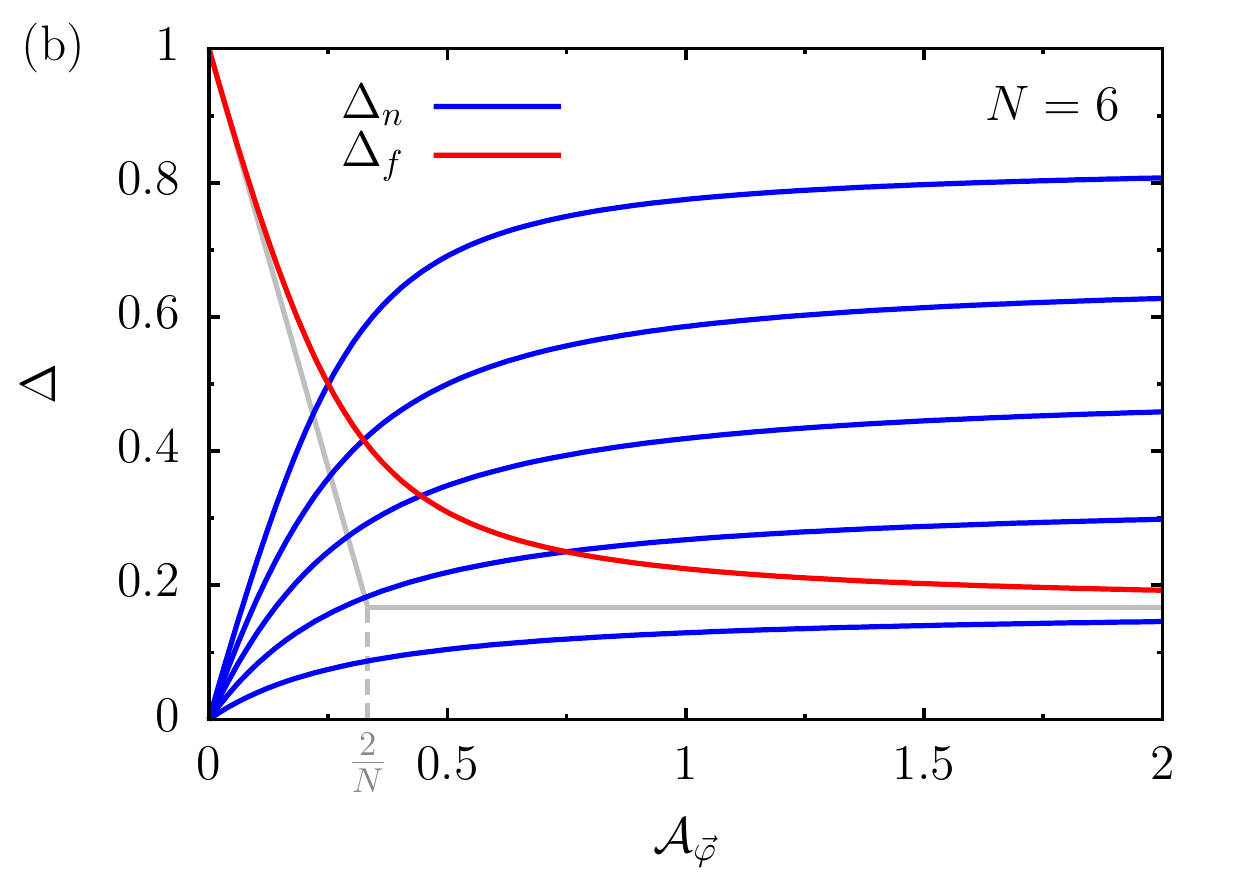}
	\caption{Spectrum $\mathcal{E}$ and gaps $\Delta$ as a function of interaction strength $\mathcal{A}_{\vec{\varphi}}\geq0$. (a) The ground state corresponds to $m_s=-\frac{N}{2}$, separated from the state $m_s=\frac{N}{2}$ by the Zeeman term equal to $1$ (in units of $\lambda\,|\mathbf{d}(\vec{\varphi})|$). Flipping a number of $n$ spins with respect to the ground state yields low- and high-energy bands $\mathcal{E}^{(2)}_{\pm,n}$. For $\mathcal{A}_{\vec{\varphi}}\gg\frac{2}{N}$, the high-energy excited bands $\mathcal{E}^{(2)}_{+,n}$ are well separated ($\propto\mathcal{A}_{\vec{\varphi}}$) from the other bands. (b) Exact results for $\Delta_n$ and $\Delta_f=1-\Delta_{N-1}$ for $N=6$. For $\mathcal{A}_{\vec{\varphi}}\ll\frac{2}{N}$, gaps $\Delta_n\approx\frac{n}{2}\,\mathcal{A}_{\vec{\varphi}}$ are described by (locally) flipping $n$ environmental spins. For $\mathcal{A}_{\vec{\varphi}}\gg\frac{2}{N}$, the gaps converge to $\Delta_n\approx\frac{n}{N}$ due to collective spin wave excitations. We have defined the crossover interaction strength $\mathcal{A}^{\text{cr}}_{\vec{\varphi}}=\frac{2}{N}$ as the intersection of the approximations $\Delta_f\approx1-\frac{N-1}{2}\,\mathcal{A}_{\vec{\varphi}}$ and $\Delta_f\approx\frac{1}{N}$.}
	\label{fig:Bandstructure}
\end{figure} 

Fig.~\hyperref[fig:Bandstructure]{\ref{fig:Bandstructure}(a)} shows the schematic structure of the instantaneous spectrum $\mathcal{E}$ as a function of ferromagnetic interaction strength $\mathcal{A}_{\vec{\varphi}}\geq0$. The ground state corresponds to $m_s=-\frac{N}{2}$, and is separated from the state $m_s=\frac{N}{2}$ by the Zeeman energy term equal to $1$ (in units of $\lambda\,|\mathbf{d}(\vec{\varphi})|$). Excited bands $\mathcal{E}^{(2)}_{\pm,n}$ are generated by flipping a number of $n$ spins relatively to the ground state. In what follows, we focus on gaps
\begin{equation}
\Delta_n(\mathcal{A}_{\vec{\varphi}})=\mathcal{E}^{(2)}_{-,n}(\mathcal{A}_{\vec{\varphi}})-\mathcal{E}^{(1)}_-(\mathcal{A}_{\vec{\varphi}}).
\label{eq:LowEnergyGaps}
\end{equation}
Exact results for $\Delta_n$ and $\Delta_f=1-\Delta_{N-1}$ are shown in Fig.~\hyperref[fig:Bandstructure]{\ref{fig:Bandstructure}(b)} for $N=6$. The underlying processes can be roughly divided into the following ranges of interaction strengths $\mathcal{A}_{\vec{\varphi}}$. For $\mathcal{A}_{\vec{\varphi}}\ll\frac{2}{N}$, the gaps $\Delta_n=\frac{n}{2}\,\mathcal{A}_{\vec{\varphi}}+\mathcal{O}(\mathcal{A}_{\vec{\varphi}}^2)$ are properly described by (locally) flipping $n$ environmental spins. By increasing the interaction strength to $\mathcal{A}_{\vec{\varphi}}\approx\frac{2}{N}$, it becomes energetically more favorable to have collective excitations similar to spin waves: a rotation of environmental spins manifest itself in a rotation of the central spin (and vice versa). This leads to low-energy bands $\mathcal{E}^{(2)}_{-,n}$ that exhibit ferromagnetic ordering $\braket{\hat{\mathbf{S}}_0\cdot\hat{\mathbf{J}}}\approx\frac{N-1}{4}$ for $\mathcal{A}_{\vec{\varphi}}\gg\frac{2}{N}$. Gaps $\Delta_n\approx\frac{n}{N}$ then arise from Zeeman splittings induced by the misalignment $\braket{\hat{S}^{z^\prime}_0}\approx-\frac{1}{2}+\frac{n}{N}$ of the central spin. High-energy bands $\mathcal{E}^{(2)}_{+,n}$ exhibit antiferromagnetic ordering $\braket{\hat{\mathbf{S}}_0\cdot\hat{\mathbf{J}}}\approx-\frac{N+1}{4}$, which is why they separate linearly ($\propto\mathcal{A}_{\vec{\varphi}}$) from the low-energy bands $\mathcal{E}^{(2)}_{-,n}$ for interaction strengths $\mathcal{A}_{\vec{\varphi}}\gg\frac{2}{N}$. From these considerations, it becomes clear why we have defined the crossover interaction strength $\mathcal{A}^{\text{cr}}_{\vec{\varphi}}=\frac{2}{N}$ as the intersection of the approximations $\Delta_f\approx1-\frac{N-1}{2}\,\mathcal{A}_{\vec{\varphi}}$ and $\Delta_f\approx\frac{1}{N}$.

In the interacting case $\mathcal{A}\neq0$, the spectrum is explicitly gapped (provided that $|\mathbf{d}(\vec{\varphi})|\neq0$) and a total Chern number $C_n=(N-2\,n)\,\nu_{gr}$ can be assigned to the synthetic energy bands of Eq.~\eqref{eq:Energies}. In fact, each flip of a spin leads to an even change in the total Chern number $\Delta C=C_{n+1}-C_{n}=-2\,\nu_{gr}$. We have introduced indices $n=0,\,1,\,\dots,\,N$ representing (again) the number of flipped spins as compared to the ferromagnetic state $m_s=-\frac{N}{2}$, and used the definition of the single-spin Chern number $\nu_{gr}$ of the main text. 

\clearpage

\section{Stability analysis for more generic couplings} \label{sec:StabilityAnalysis}
\begin{figure}
	\includegraphics[width=0.6\columnwidth]{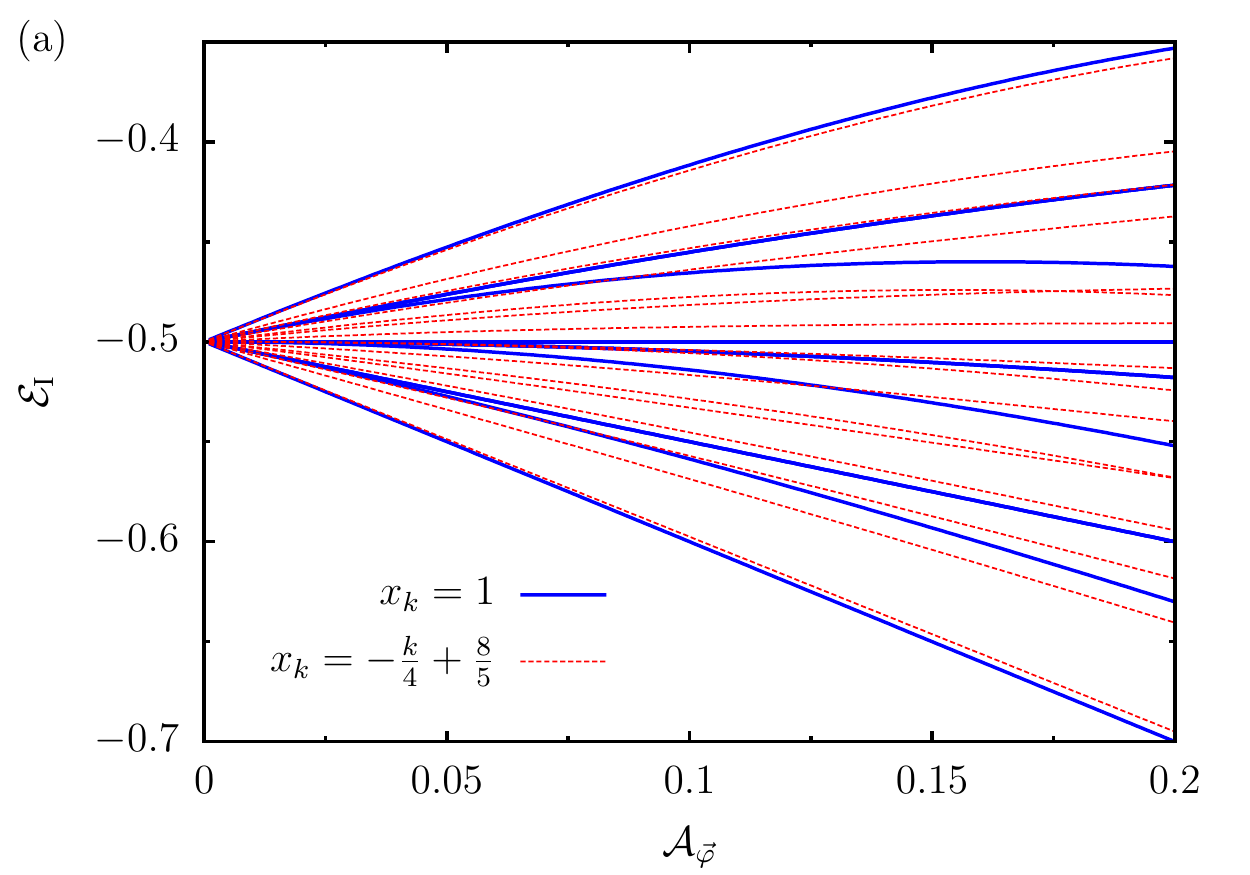}
	\includegraphics[width=0.6\columnwidth]{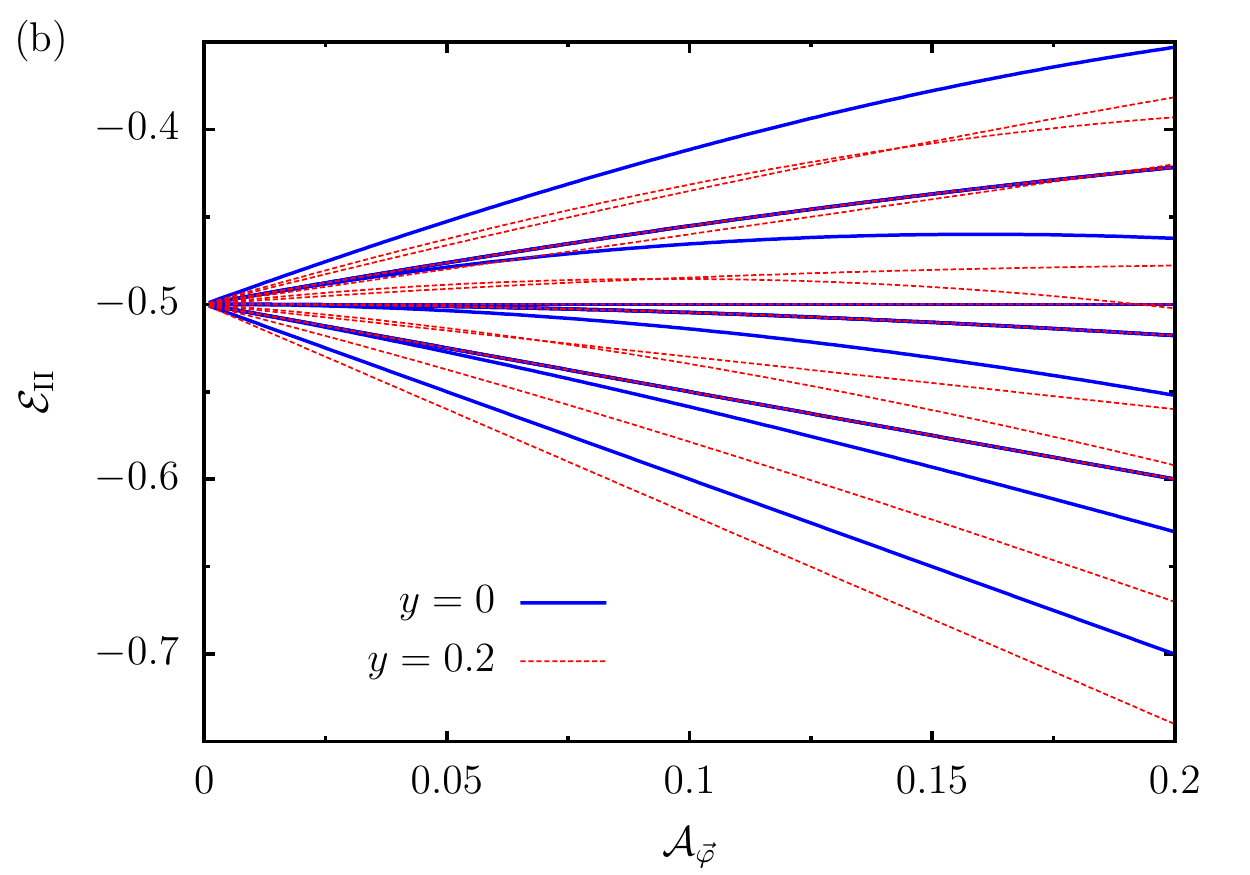}
	\caption{Spectrum $\mathcal{E}$ as a function of interaction strength $\mathcal{A}_{\vec{\varphi}}\ge0$ and a total number of spins $N=5$. For illustrative reasons, only the lower half of the synthetic energy bands is shown. (a) Results for homogeneous $x_k=1$ (solid blue lines) and inhomogeneous $x_k=-\frac{k}{4}+\frac{8}{5}$ (dashed red lines) coupling strengths $A_k=x_k\,A$. The inhomogeneity of the interaction hardly affects the ferromagnetic ground state energy band, such that its adiabatic dynamics can still be described by adiabatic perturbation theory. (b) Similar results for finite coupling strengths $I=y\,A$ between the surrounding spins with $y=0$ (solid blue lines) and $y=0.2$ (dashed red lines).}
	\label{fig:StabilityAnalysis}
\end{figure}

We explore the quasiperiodically driven CSM for more generic couplings. To this end, we consider two concrete static extensions of Eq.~\eqref{eq:CentralSpinModel} of the main text by implementing (I) inhomogeneous coupling strengths $A_k=x_k\,A$ between the central spin and the surrounding spins (see Eq.~\eqref{eq:InhomogeneousInteraction}), and (II) finite coupling strengths $I=y\,A$ between the surrounding spins (see Eq.~\eqref{eq:IntermediateInteraction}). By numerically calculating the instantaneous spectrum using exact diagonalization (see Fig.~\ref{fig:StabilityAnalysis}), we argue that the topological burning glass effect is robust to these extensions.

We start with inhomogeneous coupling strengths $A_k=x_k\,A$, yielding the Hamiltonian 
\begin{equation}
\hat{H}_{\text{I}}(\vec{\varphi}_t)=g^\ast\,\mu_B\,\mathbf{B}(\vec{\varphi}_t)\cdot\hat{\mathbf{S}}_0-A\,\sum_{k=1}^{N-1}\,x_k\,\hat{\mathbf{S}}_0\cdot\hat{\mathbf{S}}_k.
\label{eq:InhomogeneousInteraction}
\end{equation}
We have used the definitions of Eq.~\eqref{eq:CentralSpinModel} of the main text, and introduced real positive numbers $x_k\in\mathbb{R}^+$ to account for the inhomogeneity of the interaction. We analyze the energy bands $E_{\text{I}}(\vec{\varphi}_t)$ of Eq.~\eqref{eq:InhomogeneousInteraction} for fixed values of the phases $\vec{\varphi}_t$, that is at a given point of the synthetic 2D Brillouin zone (BZ). In what follows, we thus suppress the subscript $t$ from $\vec{\varphi}_t$. As in the Appendix~\ref{sec:InstantaneousSpectrum}, we perform the transformation 
\begin{equation}
\hat{H}_{\text{I}}^\prime(\vec{\varphi})=\Big(\bigotimes_{i=0}^{N-1}\,\hat{U}^\dagger(\vec{\varphi})\Big)\,\hat{H}_{\text{I}}(\vec{\varphi})\,\Big(\bigotimes_{i=0}^{N-1}\,\hat{U}(\vec{\varphi})\Big)=\lambda\,|\mathbf{d}(\vec{\varphi})|\,\Big[\hat{S}^{z^\prime}_0-\mathcal{A}_{\vec{\varphi}}\,\sum_{k=1}^{N-1}\,x_k\,\hat{\mathbf{S}}_0\cdot\hat{\mathbf{S}}_k\Big],
\label{eq:InhomogeneousInteractionRotated}
\end{equation}
where we use the definitions of Eq.~\eqref{eq:CSMRotated} and introduce the ($\vec{\varphi}$-dependent) interaction strength $\mathcal{A}_{\vec{\varphi}}=\mathcal{A}/|\mathbf{d}(\vec{\varphi})|$. Again, we assume that $|\mathbf{d}(\vec{\varphi})|\neq0$ and $\lambda>0$. Unlike the homogeneous case (see Appendix~\ref{sec:InstantaneousSpectrum}), the Hamiltonian $\hat{H}_{\text{I}}^\prime(\vec{\varphi})$ does not commute with the total spin $\hat{\mathbf{J}}^2$ of the surrounding spins, meaning that the total system can not be decomposed into a block diagonal form. Rather, the full Hilbert space must be considered, which is why we derive the dimensionless energy eigenvalues $\mathcal{E}_{\text{I}}(\mathcal{A}_{\vec{\varphi}})=\frac{E_{\text{I}}(\vec{\varphi})}{\lambda\,|\mathbf{d}(\vec{\varphi})|}$ of Eq.~\eqref{eq:InhomogeneousInteractionRotated} by exact diagonalization. Fig.~\hyperref[fig:StabilityAnalysis]{\ref{fig:StabilityAnalysis}(a)} shows the instantaneous spectrum $\mathcal{E}_{\text{I}}$ as a function of ferromagnetic interaction strength $\mathcal{A}_{\vec{\varphi}}\ge0$ and a total number of spins $N=5$ for both homogeneous $x_k=1$ (solid blue lines) and inhomogeneous $x_k=-\frac{k}{4}+\frac{8}{5}$ (dashed red lines) coupling strengths $A_k=x_k\,A$. Although some anti-crossings may arise in the excited energy bands of the spectrum, the inhomogeneity of the interaction hardly affects the ferromagnetic ground state energy band. In fact, as the latter remains energetically separated from the other bands, the adiabatic dynamics starting from the ferromagnetic ground state can still be described by adiabatic perturbation theory (cf.~Eq.~\eqref{eq:AdiabaticPerturbationTheory} of the main text). Thus, subsequent derivations of the topological burning glass effect (cf.~Eqs.~\eqref{eq:GeometricalConcepts} to \eqref{eq:ReducedDensityMatrix} of the main text) are not fundamentally affected by inhomogeneous couplings. 

Similar results can be found for finite coupling strengths $I=y\,A$ between the surrounding spins, described by the Hamiltonian 
\begin{equation}
\hat{H}_{\text{II}}(\vec{\varphi}_t)=g^\ast\,\mu_B\,\mathbf{B}(\vec{\varphi}_t)\cdot\hat{\mathbf{S}}_0-A\,\Big(\hat{\mathbf{S}}_0\cdot\hat{\mathbf{J}}+y\,\sum_{k=1}^{N-1}\,\hat{\mathbf{S}}_k\cdot\hat{\mathbf{S}}_{k+1}\Big)
\label{eq:IntermediateInteraction}
\end{equation}
with $y\in\mathbb{R}_0^+$ and periodic boundary conditions $\hat{\mathbf{S}}_N=\hat{\mathbf{S}}_1$. The dimensionless energy eigenvalues $\mathcal{E}_{\text{II}}$ associated with Eq.~\eqref{eq:IntermediateInteraction} are illustrated for $y=0$ (solid blue lines) and $y=0.2$ (dashed red lines) in Fig.~\hyperref[fig:StabilityAnalysis]{\ref{fig:StabilityAnalysis}(b)}. Again, the ferromagnetic ground state energy band remains energetically separated from the other bands. Hence, the topological burning glass effect is robust against generalizations of the CSM, e.g. for the cases of inhomogeneous couplings between central spin and surrounding spins or a finite coupling between the surrounding spins. Note that, in Fig.~\ref{fig:StabilityAnalysis}, we only show the lower half of the instantaneous spectrum for illustrative reasons, while the results for the remaining energy bands are not essential for the aforementioned discussion.

\clearpage

\section{Fluctuations around the adiabatic spin polarization of the central spin-1/2} \label{sec:DeflectionsSpinPolarization}
The reduced density matrix of Eq.~\eqref{eq:ExpectationValue} of the main text can be parametrized by vectors $\mathbf{u}^\text{dy}_{\beta}(t)$ describing the spin polarization of the central spin-1/2:
\begin{equation}
\hat{\rho}^\text{dy}_{0,\beta}(t)=\frac{1}{2}\,\Big(\mathbbm{1}+\mathbf{u}^\text{dy}_{\beta}(t)\cdot\hat{\boldsymbol{\sigma}}\Big).
\label{eq:ReducedDensityMatrixSM}
\end{equation}
A first-order expansion in $\vec{\omega}$ yields $\mathbf{u}^\text{dy}_{\beta}(t)=\mathbf{u}^\text{ad}_{\beta}(t)+\Delta\mathbf{u}_{\beta}(t)$, separated in terms of the adiabatic spin polarization $\mathbf{u}^\text{ad}_{\beta}(t)$ and the corrections $\Delta\mathbf{u}_{\beta}(t)$. Using the notation of Eq.~\eqref{eq:ReducedDensityMatrix} of the main text, the associated contributions to the reduced density matrix $\hat{\rho}^\text{dy}_{0,\beta}(t)$ can be written as
\begin{equation*}
	\hat{\rho}^\text{ad}_{0,\beta}(t)=\frac{1}{2}\,\Big(\mathbbm{1}+\mathbf{u}^\text{ad}_{\beta}(t)\cdot\hat{\boldsymbol{\sigma}}\Big),\quad \sum_{\alpha\neq\beta}\frac{\hat{T}_{\alpha\beta}(\vec{\varphi}_t)}{\Delta_{\alpha\beta}(\vec{\varphi}_t)}=\frac{1}{2}\,\Delta\mathbf{u}_{\beta}(t)\cdot\hat{\boldsymbol{\sigma}},
\end{equation*}
showing that the operators $\hat{T}_{\alpha\beta}$ induce corrections $\Delta\mathbf{u}_{\beta}(t)$ around the adiabatic spin polarization $\mathbf{u}^\text{ad}_{\beta}(t)$ of the central spin-1/2. Remarkably, these corrections contribute to the expectation value of the current operator $\hat{\text{I}}_k=\lambda\,\frac{\partial\mathbf{d}(\vec{\varphi}_t)}{\partial \varphi_k}\cdot\hat{\mathbf{S}}_0$. This can be directly seen by applying the parametrization of the reduced density matrix $\hat{\rho}^\text{dy}_{0,\beta}(t)$ of Eq.~\eqref{eq:ReducedDensityMatrixSM} to Eq.~\eqref{eq:ExpectationValue} of the main text:
\begin{equation}
\braket{\hat{\text{I}}_k}_\beta=\frac{\lambda}{2}\,\frac{\partial\mathbf{d}(\vec{\varphi}_t)}{\partial \varphi_k}\cdot\mathbf{u}^\text{ad}_{\beta}(t)+\frac{\lambda}{2}\,\frac{\partial\mathbf{d}(\vec{\varphi}_t)}{\partial \varphi_k}\cdot\Delta\mathbf{u}_{\beta}(t).
\label{eq:ExpectationValueSM}
\end{equation}
Comparing Eq.~\eqref{eq:ExpectationValueSM} with Eq.~\eqref{eq:GeometricalConcepts} of the main text, this yields  
\begin{equation}
\frac{\partial E_\beta}{\partial\varphi_k}=\frac{\lambda}{2}\,\frac{\partial\mathbf{d}(\vec{\varphi}_t)}{\partial \varphi_k}\cdot\mathbf{u}^\text{ad}_{\beta}(t),\quad \sum_{l=1}^{2}\,\omega_l\,\Omega^{(\beta)}_{kl}=\frac{\lambda}{2}\,\frac{\partial\mathbf{d}(\vec{\varphi}_t)}{\partial \varphi_k}\cdot\Delta\mathbf{u}_{\beta}(t),
\label{eq:Identities}
\end{equation}
demonstrating that the corrections $\Delta\mathbf{u}_{\beta}(t)$ include the topological information of the total system in terms of the Berry curvature $\Omega^{(\beta)}_{kl}$.

\begin{figure}[htbp]
	\includegraphics[width=\textwidth]{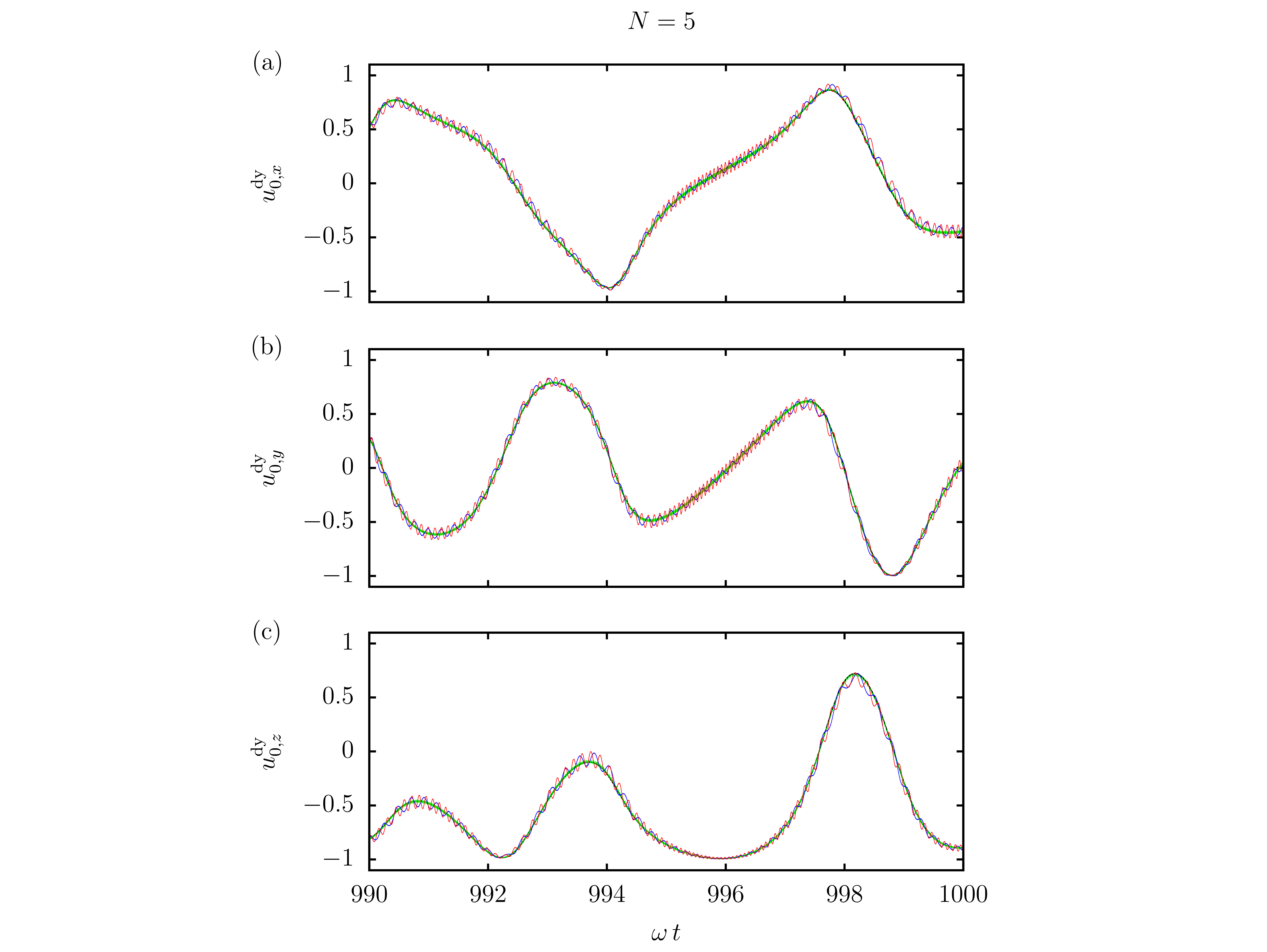}
	\caption{Spin polarization $\mathbf{u}^\text{dy}_{0}=(u^\text{dy}_{0,x},u^\text{dy}_{0,y},u^\text{dy}_{0,z})$ as a function of time $\omega\,t$ for different interaction strengths $\mathcal{A}=0.0$ (green), $\mathcal{A}=0.5$ (blue), $\mathcal{A}=4.0$ (red). We have selected a total number of spins $N=5$ and a frequency $\omega/\lambda=5\cdot10^{-3}$. The adiabatic spin polarization $\mathbf{u}^\text{ad}_{0}(t)=-\tilde{\mathbf{d}}(\vec{\varphi}_t)$ is represented as a solid black line. Note that the green line only slightly deviates from the black line.}
	\label{fig:SpinPolarization_N5}
\end{figure}

In Fig.~\ref{fig:SpinPolarization_N5}, we illustrate the numerically exact simulations of the spin polarization $\mathbf{u}^\text{dy}_{0}(t)=(u^\text{dy}_{0,x}(t),u^\text{dy}_{0,y}(t),u^\text{dy}_{0,z}(t))$ as a function of time $\omega\,t$ for different ferromagnetic interaction strengths $\mathcal{A}=0.0$ (green), $\mathcal{A}=0.5$ (blue), $\mathcal{A}=4.0$ (red). We initialize the system at $t_0=0$ in the ferromagnetic ground state, selecting a total number of spins $N=5$ and a frequency $\omega/\lambda=5\cdot10^{-3}$. The remaining parameters are chosen as in Fig.~\ref{fig:PhaseDiagram} of the main text. The adiabatic spin polarization $\mathbf{u}^\text{ad}_{0}(t)=-\tilde{\mathbf{d}}(\vec{\varphi}_t)$ is represented as a solid black line, where we have used the definition $\tilde{\mathbf{d}}(\vec{\varphi})=\frac{\mathbf{d}(\vec{\varphi})}{|\mathbf{d}(\vec{\varphi})|}$ of the main text. In the interacting case $\mathcal{A}\neq0$, the spin polarization $\mathbf{u}^\text{dy}_{0}(t)$ differs from the single spin case ($\mathcal{A}=0$), indicating that the non-unitary dynamics of the central spin is crucially affected by the geometrical properties of the total system. Note that this information is contained in the enhanced fluctuations (compared to the single spin case) around the black line in Fig.~\ref{fig:SpinPolarization_N5}.

\clearpage

\section{Nonadiabatic breakdown of the strong coupling regime} \label{sec:BreakdownStrongCouplingRegime}
In what follows, we refer to the topological band structure of Fig.~\ref{fig:Bandstructure}. Starting from the ferromagnetic ground state energy band $\mathcal{E}^{(1)}_-$, the nonadiabatic breakdown of the strong coupling regime occurs in two main stages.

In stage (I), the quantum state shows equal time-averaged overlaps with the instantaneous eigenstates of the energy bands $\mathcal{E}^{(1)}_-$ and $\mathcal{E}^{(2)}_{-,n}$, resulting in an intermediate dynamical quantum phase with an ensemble-averaged pumping rate
\begin{equation}
\bar{P}^{12}=-\bar{P}^{21}=\frac{1}{N}\sum_{n=0}^{N-1}P_n^{12}=\frac{\nu_{gr}}{2\,\pi}\,\omega_1\,\omega_2.
\label{eq:EnsembleAveragedPower}
\end{equation}
Each occupied energy band contributes by a time-averaged pumping rate $P_n^{12}=-P_n^{21}=\frac{C_n}{2\,\pi}\,\omega_1\,\omega_2$ determined by the total Chern number $C_n=(N-2\,n)\,\nu_{gr}$. The ensemble-averaged pumping rate of Eq.~\eqref{eq:EnsembleAveragedPower} can be observed for interaction strengths $\mathcal{A}_{\vec{\varphi}}>0$ that induce low-energy gaps $\Delta_1\ll\Delta_f$ (see Fig.~\ref{fig:Bandstructure}): once the dynamics leads to nonadiabatic excitation processes to the first excited energy band $\mathcal{E}^{(2)}_{-,1}$, subsequent energy bands $\mathcal{E}^{(2)}_{-,n}$ are gradually occupied as gaps $\Delta_1$ and $\Delta_{n+1}-\Delta_n$ are of similar size. Populations in the instantaneous ground $\mathcal{E}^{(1)}_-$ and low-energy bands $\mathcal{E}^{(2)}_{-,n}$ then become equal within an intermediate (pre)thermal regime, while transitions to the energy band $\mathcal{E}^{(1)}_+$ are suppressed. For $\mathcal{A}_{\vec{\varphi}}\gg\frac{2}{N}$, however, all low-energy gaps $\Delta_1\approx\Delta_f\approx\Delta_{n+1}-\Delta_n$ have approximately the same size. (Pre)thermalization then immediately leads to a featureless quantum phase (II) with zero time-averaged pumping rate $P=0$, as the total Chern numbers sum to $\sum_{n=0}^{N}C_n=\sum_{n=1}^{N-1}C_n=0$. 

Stage (II) divides into two phases. (IIa) In addition to the overlaps of the intermediate phase, also the energy band $\mathcal{E}^{(1)}_+$ is equally occupied on time-average. (IIb) Within the block $j=\frac{N-1}{2}$, an ergodic regime arises in which the quantum state additionally exhibits equal time-averaged overlaps with the instantaneous eigenstates of the high-energy bands $\mathcal{E}^{(2)}_{+,n}$. Stages (I)-(II) can be considered as (pre)thermalization processes. 

\begin{figure*}[ht!]
	\centering
	\includegraphics[width=\textwidth]{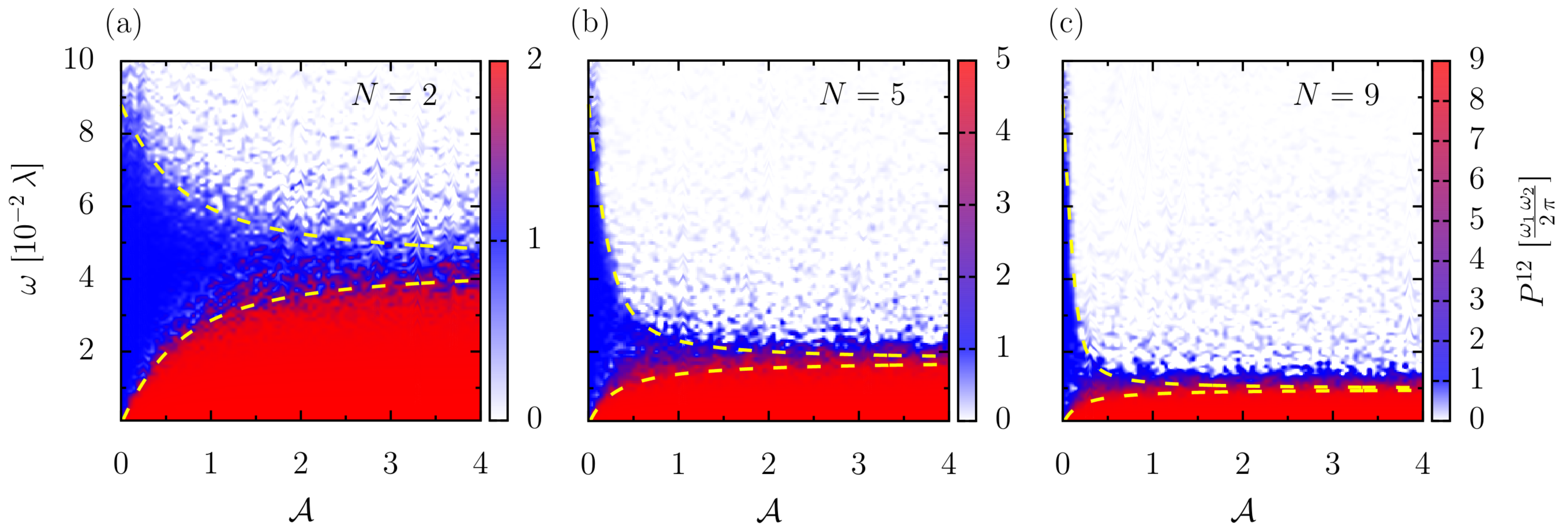}	
	\caption{Nonequilibrium phase diagrams as a function of interaction strength $\mathcal{A}>0$ and frequencies $\omega_1=\omega$, $\omega_2=\gamma\,\omega$ for different total spin numbers $N$. Provided the adiabatic dynamics is confined to the ferromagnetic ground state, the time-averaged pumping rate $P^{12}$ becomes $N$-fold enhanced by the collective motion of the surrounding spins (red regimes). Nonadiabatic excitation processes can result in dynamical transitions to intermediate/featureless quantum phases with pumping rates $\bar{P}^{12}=\frac{\nu_{gr}}{2\,\pi}\,\omega_1\,\omega_2$ (blue regimes)/$P^{12}=0$ (white regimes). The critical frequencies $\omega_c$ scale with the associated gaps (see Fig.~\ref{fig:Bandstructure}), roughly illustrated by phase boundaries $\omega_c=0.11\,\Delta_{1/f}^{\text{min}}$ (dashed yellow lines).}
	\label{fig:PhaseDiagramSM}
\end{figure*}

In Fig.~\ref{fig:PhaseDiagramSM}, the time-averaged pumping rate $P^{12}$ is shown as a function of ferromagnetic interaction strength $\mathcal{A}>0$ and frequency $\omega$ for different total numbers of spins $N$. The techniques and remaining parameters are chosen as in Fig.~\ref{fig:PhaseDiagram} of the main text. The topological frequency conversion occurs in accordance with the previous discussion. The phase boundaries of the dynamical transitions are not sharp, as they originate from nonanalytical excitation processes. Nevertheless, the numerics suggest that the leading processes are approximately covered by critical frequencies $\omega_c\propto\Delta_{1/f}^{\text{min}}$, illustrated by assuming a proportionality constant $K_T=0.11$ in Fig.~\ref{fig:PhaseDiagramSM} (dashed yellow lines). We have evaluated the gaps $\Delta_{1/f}$ of Eq.~\eqref{eq:LowEnergyGaps} in the origin ($\vec{\varphi}=0$) of the synthetic 2D Brillouin zone (BZ), obtaining the minimum band gaps 
\begin{equation*}
	\Delta_{1/f}^{\text{min}}(\mathcal{A})=\lambda\,|\mathbf{d}(0)|\,\Delta_{1/f}\Big(\frac{\mathcal{A}}{|\mathbf{d}(0)|}\Big)
\end{equation*} 
for the mass parameter $M=1.2$. The approach for the critical frequencies $\omega_c$ becomes less accurate with larger total spin numbers $N$, especially evident in Figs.~\hyperref[fig:PhaseDiagramSM]{\ref{fig:PhaseDiagramSM}(b)} and \hyperref[fig:PhaseDiagramSM]{\ref{fig:PhaseDiagramSM}(c)}. This, however, can be explained by the time-scales required for (pre)thermalization into the nonequilibrium quantum phases: with a higher number of energy bands (controlled by a larger total number of spins $N$), transition phases with pumping rates beyond the integer-quantized values become more pronounced. This transient effect, in turn, leads to a broadening of the phase boundaries of the dynamical transitions, which slightly distorts the approximation $\omega_c\propto\Delta_{1/f}^{\text{min}}$. Nevertheless, the fundamental scaling behavior of the phase diagrams of Fig.~\ref{fig:PhaseDiagramSM} can be analyzed (see also Fig.~\ref{fig:Bandstructure}): both the crossover interaction strength $\mathcal{A}^{\text{cr}}=\frac{2\,|\mathbf{d}(0)|}{N}=\frac{1.6}{N}$ and the minimum band gaps $\Delta_{1/f}^{\text{min}}/\lambda\approx\frac{0.8}{N}$ scale with the inverse of $N$ for a larger number of environmental spins. We have used that $|\mathbf{d}(0)|=0.8$. 

\clearpage

\section{Results for antiferromagnetic interaction strengths} \label{sec:AntiferromagneticInteraction}
\begin{figure}
	\includegraphics[width=0.6\columnwidth]{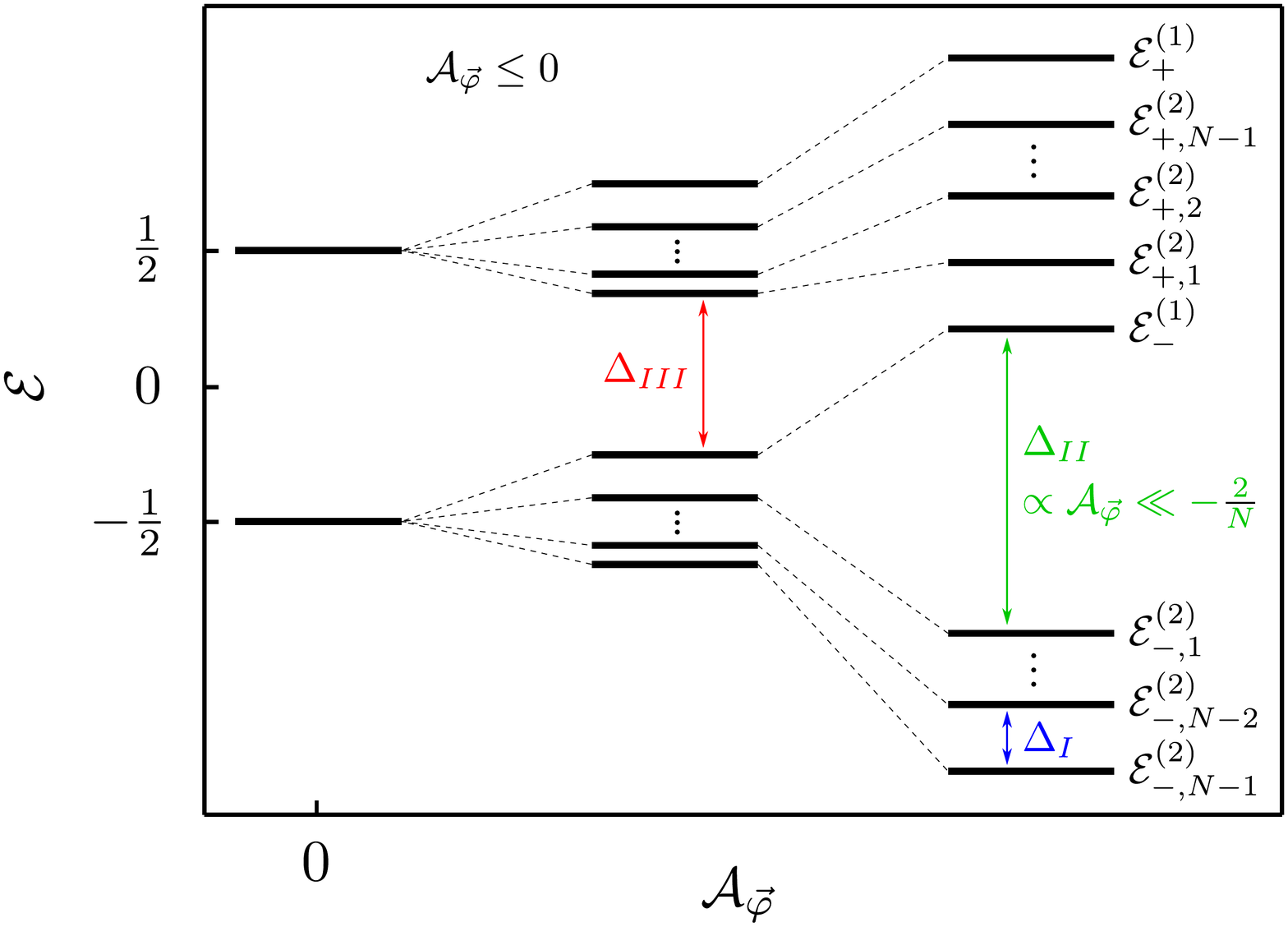}
	\caption{Spectrum $\mathcal{E}$ as a function of interaction strength $\mathcal{A}_{\vec{\varphi}}\leq0$. The ground state energy band corresponds to $\mathcal{E}^{(2)}_{-,N-1}$, followed by low-energy bands $\mathcal{E}^{(2)}_{-,n}$ with decreasing number $n$. High-energy excited bands $\mathcal{E}^{(2)}_{+,n}$ are located between states $m_s=-\frac{N}{2}$ and $m_s=\frac{N}{2}$, the latter separated by the Zeeman energy term equal to $1$ (in units of $\lambda\,|\mathbf{d}(\vec{\varphi})|$). For $\mathcal{A}_{\vec{\varphi}}\ll-\frac{2}{N}$, the low-energy bands $\mathcal{E}^{(2)}_{-,n}$ are well separated ($\propto\mathcal{A}_{\vec{\varphi}}$) from the other bands. Gaps $\Delta_{I-III}$ relevant for the nonequilibrium evolution (see Fig.~\ref{fig:PhaseDiagram_AF}) are schematically illustrated, while gaps $\Delta_{II}=\Delta_{III}$ are exactly the same for an interaction strength $\mathcal{A}_{\vec{\varphi}}=-\frac{2}{N}$.}
	\label{fig:Bandstructure_AF}
\end{figure}

\begin{figure}
	\includegraphics[width=\textwidth]{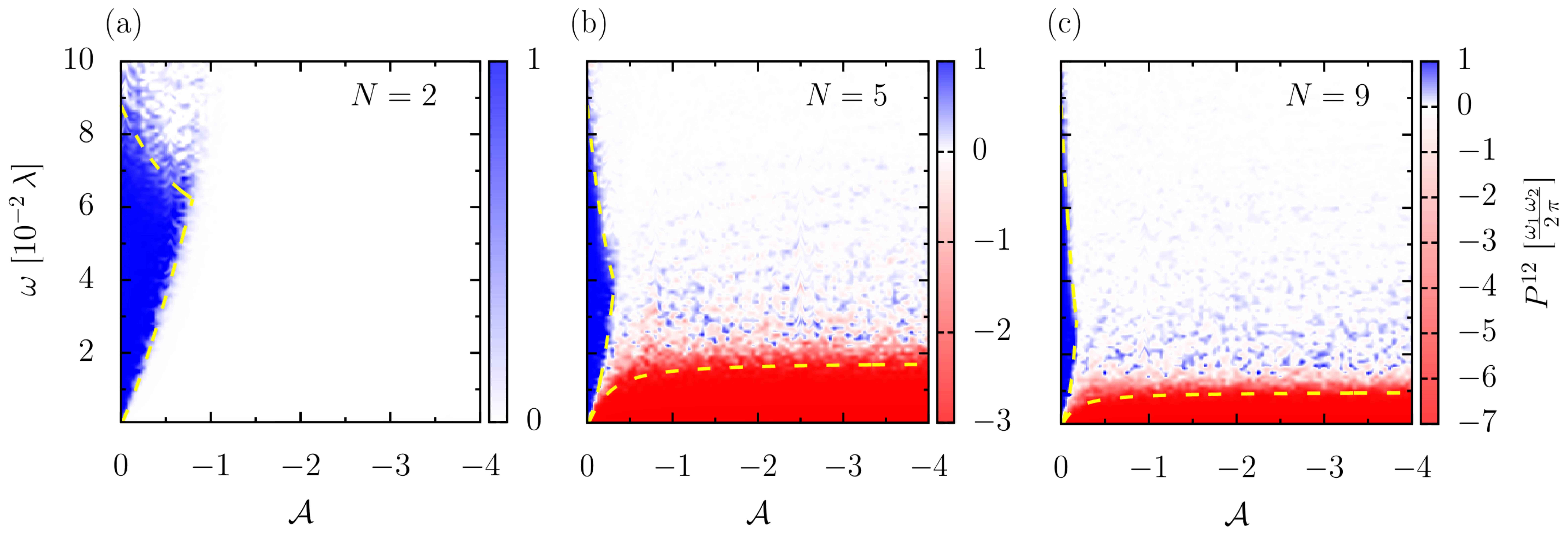}
	\caption{Nonequilibrium phase diagrams as a function of interaction strength $\mathcal{A}<0$ and frequencies $\omega_1=\omega$, $\omega_2=\gamma\,\omega$ for different total spin numbers $N$. The system is initially prepared in the antiferromagnetic ground state energy band $\mathcal{E}^{(2)}_{-,N-1}$. Numerical methods and associated parameters are chosen as in Fig.~\ref{fig:PhaseDiagram} of the main text. Provided the adiabatic dynamics is confined to the energy band $\mathcal{E}^{(2)}_{-,N-1}$, the time-averaged pumping rate $P^{12}$ is proportional to the total Chern number $C_{N-1}=(2-N)\,\nu_{gr}$ (red regimes). Nonadiabatic excitation processes can in turn lead to dynamical transitions to intermediate/featureless quantum phases with pumping rates $\bar{P}^{12}=\frac{\nu_{gr}}{2\,\pi}\,\omega_1\,\omega_2$ (blue regimes)/$P^{12}=0$ (white regimes). The critical frequencies $\omega_c$ scale with the associated gaps (see Fig.~\ref{fig:Bandstructure_AF}), roughly illustrated by phase boundaries $\omega_c=0.11\,\Delta_{I-III}^{\text{min}}$ (dashed yellow lines). }
	\label{fig:PhaseDiagram_AF}
\end{figure}

We examine the topological band structure and the nonequilibrium phase diagrams of the quasiperiodically driven CSM for an antiferromagnetic interaction strength $\mathcal{A}<0$. Analogous to the ferromagnetic case (see Fig.~\hyperref[fig:Bandstructure]{\ref{fig:Bandstructure}(a)}), the energy bands \eqref{eq:Energies} can be energetically ordered for an antiferromagnetic interaction strength. Fig.~\ref{fig:Bandstructure_AF} shows the schematic structure of the instantaneous spectrum $\mathcal{E}$ as a function of antiferromagnetic interaction strength $\mathcal{A}_{\vec{\varphi}}\leq0$. The antiferromagnetic ground state energy band corresponds to $\mathcal{E}^{(2)}_{-,N-1}$, followed by low-energy bands $\mathcal{E}^{(2)}_{-,n}$ with gradually decreasing number $n$. Note that the number $n=1,\,2,\,\cdots,\,N-1$ of flipped spins is again defined with respect to the ferromagnetic state $m_s=-\frac{N}{2}$. The high-energy bands $\mathcal{E}^{(2)}_{+,n}$ are located between the ferromagnetic states $m_s=-\frac{N}{2}$ and $m_s=\frac{N}{2}$, with the latter again separated by the Zeeman energy term equal to $1$ (in units of $\lambda\,|\mathbf{d}(\vec{\varphi})|$). We restrict ourselves to gaps
\begin{equation}
\Delta_I(\mathcal{A}_{\vec{\varphi}})=\mathcal{E}^{(2)}_{-,N-2}(\mathcal{A}_{\vec{\varphi}})-\mathcal{E}^{(2)}_{-,N-1}(\mathcal{A}_{\vec{\varphi}}), \quad \Delta_{II}=\mathcal{E}^{(1)}_-(\mathcal{A}_{\vec{\varphi}})-\mathcal{E}^{(2)}_{-,1}(\mathcal{A}_{\vec{\varphi}}), \quad \Delta_{III}(\mathcal{A}_{\vec{\varphi}})=\mathcal{E}^{(2)}_{+,1}(\mathcal{A}_{\vec{\varphi}})-\mathcal{E}^{(1)}_-(\mathcal{A}_{\vec{\varphi}}),
\label{eq:Gaps_AF}
\end{equation}
which are relevant for the formation of the dynamical topological quantum phases in the antiferromagnetic case. For $\mathcal{A}_{\vec{\varphi}}\ll-\frac{2}{N}$, the gaps converge to $\Delta_I\approx\Delta_{III}\approx\frac{1}{N}$ due to collective spin wave excitations, while $\Delta_{II}\propto\mathcal{A}_{\vec{\varphi}}$ indicates that the energy difference between low- $\mathcal{E}^{(2)}_{-,n}$ and high-lying bands $\mathcal{E}^{(2)}_{+,n}$ goes again linearly with the interaction strength $\mathcal{A}_{\vec{\varphi}}$. The corresponding nonequilibrium phase diagrams as a function of antiferromagnetic interaction strength $\mathcal{A}<0$ and frequency $\omega$ are shown in Fig.~\ref{fig:PhaseDiagram_AF} for different total numbers of spins $N$. The numerical methods and associated parameters are chosen as in Fig.~\ref{fig:PhaseDiagram} of the main text, except that the system is initially prepared in the antiferromagnetic ground state energy band $\mathcal{E}^{(2)}_{-,N-1}$. Provided the adiabatic dynamics is confined to the energy band $\mathcal{E}^{(2)}_{-,N-1}$, the time-averaged pumping rate $P^{12}$ is proportional to the total Chern number $C_{N-1}=(2-N)\,\nu_{gr}$ (red regimes). An ensemble-averaged pumping rate $\bar{P}^{12}=\frac{\nu_{gr}}{2\,\pi}\,\omega_1\,\omega_2$ (see Eq.~\eqref{eq:EnsembleAveragedPower}) identical to the noninteracting case (blue regimes) can be observed for interaction strengths $\mathcal{A}_{\vec{\varphi}}<0$ that induce energy gaps $\Delta_{II}\ll\Delta_{III}$ (see Fig.~\ref{fig:Bandstructure_AF}): populations in the low-energy bands $\mathcal{E}^{(2)}_{-,n}$ and the ferromagnetic energy band $\mathcal{E}^{(1)}_-$ become equal within an intermediate (pre)thermal regime, while transitions to the high-energy band $\mathcal{E}^{(2)}_{+,1}$ are suppressed. For $\Delta_{II}\gtrapprox\Delta_{III}$, (pre)thermalization immediately leads to a featureless quantum phase with zero time-averaged pumping rate $P^{12}=0$ (white regimes), as the total Chern numbers sum to $\sum_{n=1}^{N-1}\,C_n=\sum_{n=0}^{N}\,C_n=0$. Gaps $\Delta_{II}=\Delta_{III}$ are exactly the same for $\mathcal{A}_{\vec{\varphi}}=-\frac{2}{N}$, which makes this especially relevant for interaction strengths $\mathcal{A}_{\vec{\varphi}}\ll-\frac{2}{N}$. However, (pre)thermalization into a featureless quantum state can also be observed in general for interaction strengths $\mathcal{A}_{\vec{\varphi}}<0$ that induce low-energy gaps $\Delta_I\ll\Delta_{II}$, completely independent of the gaps $\Delta_{III}$. As in the Appendix~\ref{sec:BreakdownStrongCouplingRegime}, the exact dynamics suggests that the leading nonadiabatic excitation processes are approximately covered by critical frequencies $\omega_c\propto\Delta_{I-III}^{\text{min}}$, illustrated by assuming a proportionality constant $K=0.11$ in Fig.~\ref{fig:PhaseDiagram_AF} (dashed yellow lines). Again, we have evaluated the gaps $\Delta_{I-III}$ of Eq.~\eqref{eq:Gaps_AF} in the origin ($\vec{\varphi}=0$) of the synthetic 2D Brillouin zone (BZ), obtaining the minimum band gaps
\begin{equation*}
	\Delta_{I-III}^{\text{min}}(\mathcal{A})=\lambda\,|\mathbf{d}(0)|\,\Delta_{I-III}\Big(\frac{\mathcal{A}}{|\mathbf{d}(0)|}\Big).
\end{equation*}        

\begin{thebibliography}{53}%
	\makeatletter
	\providecommand \@ifxundefined [1]{%
		\@ifx{#1\undefined}
	}%
	\providecommand \@ifnum [1]{%
		\ifnum #1\expandafter \@firstoftwo
		\else \expandafter \@secondoftwo
		\fi
	}%
	\providecommand \@ifx [1]{%
		\ifx #1\expandafter \@firstoftwo
		\else \expandafter \@secondoftwo
		\fi
	}%
	\providecommand \natexlab [1]{#1}%
	\providecommand \enquote  [1]{``#1''}%
	\providecommand \bibnamefont  [1]{#1}%
	\providecommand \bibfnamefont [1]{#1}%
	\providecommand \citenamefont [1]{#1}%
	\providecommand \href@noop [0]{\@secondoftwo}%
	\providecommand \href [0]{\begingroup \@sanitize@url \@href}%
	\providecommand \@href[1]{\@@startlink{#1}\@@href}%
	\providecommand \@@href[1]{\endgroup#1\@@endlink}%
	\providecommand \@sanitize@url [0]{\catcode `\\12\catcode `\$12\catcode
		`\&12\catcode `\#12\catcode `\^12\catcode `\_12\catcode `\%12\relax}%
	\providecommand \@@startlink[1]{}%
	\providecommand \@@endlink[0]{}%
	\providecommand \url  [0]{\begingroup\@sanitize@url \@url }%
	\providecommand \@url [1]{\endgroup\@href {#1}{\urlprefix }}%
	\providecommand \urlprefix  [0]{URL }%
	\providecommand \Eprint [0]{\href }%
	\providecommand \doibase [0]{http://dx.doi.org/}%
	\providecommand \selectlanguage [0]{\@gobble}%
	\providecommand \bibinfo  [0]{\@secondoftwo}%
	\providecommand \bibfield  [0]{\@secondoftwo}%
	\providecommand \translation [1]{[#1]}%
	\providecommand \BibitemOpen [0]{}%
	\providecommand \bibitemStop [0]{}%
	\providecommand \bibitemNoStop [0]{.\EOS\space}%
	\providecommand \EOS [0]{\spacefactor3000\relax}%
	\providecommand \BibitemShut  [1]{\csname bibitem#1\endcsname}%
	\let\auto@bib@innerbib\@empty
	%</preamble>
	\bibitem [{\citenamefont {Thouless}(1998)}]{Thouless_1998}%
	\BibitemOpen
	\bibfield  {author} {\bibinfo {author} {\bibfnamefont {D.~J.}\ \bibnamefont
			{Thouless}},\ }\href@noop {} {\emph {\bibinfo {title} {Topological Quantum
				Numbers in Nonrelativistic Physics}}}\ (\bibinfo  {publisher} {World
		Scientific},\ \bibinfo {address} {Singapore},\ \bibinfo {year}
	{1998})\BibitemShut {NoStop}%
	\bibitem [{\citenamefont {Xiao}\ \emph {et~al.}(2010)\citenamefont {Xiao},
		\citenamefont {Chang},\ and\ \citenamefont {Niu}}]{Xiao2010}%
	\BibitemOpen
	\bibfield  {author} {\bibinfo {author} {\bibfnamefont {D.}~\bibnamefont
			{Xiao}}, \bibinfo {author} {\bibfnamefont {M.-C.}\ \bibnamefont {Chang}}, \
		and\ \bibinfo {author} {\bibfnamefont {Q.}~\bibnamefont {Niu}},\ }\href
	{\doibase 10.1103/RevModPhys.82.1959} {\bibfield  {journal} {\bibinfo
			{journal} {Rev. Mod. Phys.}\ }\textbf {\bibinfo {volume} {82}},\ \bibinfo
		{pages} {1959} (\bibinfo {year} {2010})}\BibitemShut {NoStop}%
	\bibitem [{\citenamefont {Wen}(2017)}]{Wen_2017}%
	\BibitemOpen
	\bibfield  {author} {\bibinfo {author} {\bibfnamefont {X.-G.}\ \bibnamefont
			{Wen}},\ }\href {\doibase 10.1103/RevModPhys.89.041004} {\bibfield  {journal}
		{\bibinfo  {journal} {Rev. Mod. Phys.}\ }\textbf {\bibinfo {volume} {89}},\
		\bibinfo {pages} {041004} (\bibinfo {year} {2017})}\BibitemShut {NoStop}%
	\bibitem [{\citenamefont {Cayssol}\ and\ \citenamefont
		{Fuchs}(2021)}]{Cayssol_2021}%
	\BibitemOpen
	\bibfield  {author} {\bibinfo {author} {\bibfnamefont {J.}~\bibnamefont
			{Cayssol}}\ and\ \bibinfo {author} {\bibfnamefont {J.~N.}\ \bibnamefont
			{Fuchs}},\ }\href {\doibase 10.1088/2515-7639/abf0b5} {\bibfield  {journal}
		{\bibinfo  {journal} {J. Phys.: Materials}\ }\textbf {\bibinfo {volume}
			{4}},\ \bibinfo {pages} {034007} (\bibinfo {year} {2021})}\BibitemShut
	{NoStop}%
	\bibitem [{\citenamefont {Klitzing}\ \emph {et~al.}(1980)\citenamefont
		{Klitzing}, \citenamefont {Dorda},\ and\ \citenamefont
		{Pepper}}]{Klitzing_1980}%
	\BibitemOpen
	\bibfield  {author} {\bibinfo {author} {\bibfnamefont {K.~v.}\ \bibnamefont
			{Klitzing}}, \bibinfo {author} {\bibfnamefont {G.}~\bibnamefont {Dorda}}, \
		and\ \bibinfo {author} {\bibfnamefont {M.}~\bibnamefont {Pepper}},\ }\href
	{\doibase 10.1103/PhysRevLett.45.494} {\bibfield  {journal} {\bibinfo
			{journal} {Phys. Rev. Lett.}\ }\textbf {\bibinfo {volume} {45}},\ \bibinfo
		{pages} {494} (\bibinfo {year} {1980})}\BibitemShut {NoStop}%
	\bibitem [{\citenamefont {Prange}\ and\ \citenamefont
		{Girvin}(1990)}]{Prange_1990}%
	\BibitemOpen
	\bibfield  {author} {\bibinfo {author} {\bibfnamefont {R.~E.}\ \bibnamefont
			{Prange}}\ and\ \bibinfo {author} {\bibfnamefont {S.~M.}\ \bibnamefont
			{Girvin}},\ }\href@noop {} {\emph {\bibinfo {title} {The Quantum Hall
				Effect}}},\ \bibinfo {edition} {2nd}\ ed.\ (\bibinfo  {publisher}
	{Springer},\ \bibinfo {address} {New York},\ \bibinfo {year}
	{1990})\BibitemShut {NoStop}%
	\bibitem [{\citenamefont {Laughlin}(1981)}]{Laughlin_1981}%
	\BibitemOpen
	\bibfield  {author} {\bibinfo {author} {\bibfnamefont {R.~B.}\ \bibnamefont
			{Laughlin}},\ }\href {\doibase 10.1103/PhysRevB.23.5632} {\bibfield
		{journal} {\bibinfo  {journal} {Phys. Rev. B}\ }\textbf {\bibinfo {volume}
			{23}},\ \bibinfo {pages} {5632} (\bibinfo {year} {1981})}\BibitemShut
	{NoStop}%
	\bibitem [{\citenamefont {Thouless}\ \emph {et~al.}(1982)\citenamefont
		{Thouless}, \citenamefont {Kohmoto}, \citenamefont {Nightingale},\ and\
		\citenamefont {den Nijs}}]{Thouless_1982}%
	\BibitemOpen
	\bibfield  {author} {\bibinfo {author} {\bibfnamefont {D.~J.}\ \bibnamefont
			{Thouless}}, \bibinfo {author} {\bibfnamefont {M.}~\bibnamefont {Kohmoto}},
		\bibinfo {author} {\bibfnamefont {M.~P.}\ \bibnamefont {Nightingale}}, \ and\
		\bibinfo {author} {\bibfnamefont {M.}~\bibnamefont {den Nijs}},\ }\href
	{\doibase 10.1103/PhysRevLett.49.405} {\bibfield  {journal} {\bibinfo
			{journal} {Phys. Rev. Lett.}\ }\textbf {\bibinfo {volume} {49}},\ \bibinfo
		{pages} {405} (\bibinfo {year} {1982})}\BibitemShut {NoStop}%
	\bibitem [{\citenamefont {Simon}(1983)}]{Simon_1983}%
	\BibitemOpen
	\bibfield  {author} {\bibinfo {author} {\bibfnamefont {B.}~\bibnamefont
			{Simon}},\ }\href {\doibase 10.1103/PhysRevLett.51.2167} {\bibfield
		{journal} {\bibinfo  {journal} {Phys. Rev. Lett.}\ }\textbf {\bibinfo
			{volume} {51}},\ \bibinfo {pages} {2167} (\bibinfo {year}
		{1983})}\BibitemShut {NoStop}%
	\bibitem [{\citenamefont {Haldane}(1988)}]{Haldane1988}%
	\BibitemOpen
	\bibfield  {author} {\bibinfo {author} {\bibfnamefont {F.~D.~M.}\
			\bibnamefont {Haldane}},\ }\href {\doibase 10.1103/PhysRevLett.61.2015}
	{\bibfield  {journal} {\bibinfo  {journal} {Phys. Rev. Lett.}\ }\textbf
		{\bibinfo {volume} {61}},\ \bibinfo {pages} {2015} (\bibinfo {year}
		{1988})}\BibitemShut {NoStop}%
	\bibitem [{\citenamefont {Kubo}(1957)}]{Kubo_1957}%
	\BibitemOpen
	\bibfield  {author} {\bibinfo {author} {\bibfnamefont {R.}~\bibnamefont
			{Kubo}},\ }\href {\doibase 10.1143/JPSJ.12.570} {\bibfield  {journal}
		{\bibinfo  {journal} {J. Phys. Soc. Jpn.}\ }\textbf {\bibinfo {volume}
			{12}},\ \bibinfo {pages} {570} (\bibinfo {year} {1957})}\BibitemShut
	{NoStop}%
	\bibitem [{\citenamefont {Mahan}(2000)}]{Mahan_2000}%
	\BibitemOpen
	\bibfield  {author} {\bibinfo {author} {\bibfnamefont {G.~D.}\ \bibnamefont
			{Mahan}},\ }\href@noop {} {\emph {\bibinfo {title} {Many-Particle
				Physics}}},\ \bibinfo {edition} {3rd}\ ed.\ (\bibinfo  {publisher} {Kluwer},\
	\bibinfo {address} {Boston},\ \bibinfo {year} {2000})\BibitemShut {NoStop}%
	\bibitem [{\citenamefont {Chern}(1946)}]{Chern_1946}%
	\BibitemOpen
	\bibfield  {author} {\bibinfo {author} {\bibfnamefont {S.~S.}\ \bibnamefont
			{Chern}},\ }\href {\doibase 10.2307/1969037} {\bibfield  {journal} {\bibinfo
			{journal} {Annals of Mathematics}\ }\textbf {\bibinfo {volume} {47}},\
		\bibinfo {pages} {85} (\bibinfo {year} {1946})}\BibitemShut {NoStop}%
	\bibitem [{\citenamefont {Qi}\ \emph {et~al.}(2008)\citenamefont {Qi},
		\citenamefont {Hughes},\ and\ \citenamefont {Zhang}}]{Qi_2008}%
	\BibitemOpen
	\bibfield  {author} {\bibinfo {author} {\bibfnamefont {X.-L.}\ \bibnamefont
			{Qi}}, \bibinfo {author} {\bibfnamefont {T.~L.}\ \bibnamefont {Hughes}}, \
		and\ \bibinfo {author} {\bibfnamefont {S.-C.}\ \bibnamefont {Zhang}},\ }\href
	{\doibase 10.1103/PhysRevB.78.195424} {\bibfield  {journal} {\bibinfo
			{journal} {Phys. Rev. B}\ }\textbf {\bibinfo {volume} {78}},\ \bibinfo
		{pages} {195424} (\bibinfo {year} {2008})}\BibitemShut {NoStop}%
	\bibitem [{\citenamefont {Hasan}\ and\ \citenamefont {Kane}(2010)}]{Hasan2010}%
	\BibitemOpen
	\bibfield  {author} {\bibinfo {author} {\bibfnamefont {M.~Z.}\ \bibnamefont
			{Hasan}}\ and\ \bibinfo {author} {\bibfnamefont {C.~L.}\ \bibnamefont
			{Kane}},\ }\href {\doibase 10.1103/RevModPhys.82.3045} {\bibfield  {journal}
		{\bibinfo  {journal} {Rev. Mod. Phys.}\ }\textbf {\bibinfo {volume} {82}},\
		\bibinfo {pages} {3045} (\bibinfo {year} {2010})}\BibitemShut {NoStop}%
	\bibitem [{\citenamefont {Qi}\ and\ \citenamefont {Zhang}(2011)}]{Qi_2011}%
	\BibitemOpen
	\bibfield  {author} {\bibinfo {author} {\bibfnamefont {X.-L.}\ \bibnamefont
			{Qi}}\ and\ \bibinfo {author} {\bibfnamefont {S.-C.}\ \bibnamefont {Zhang}},\
	}\href {\doibase 10.1103/RevModPhys.83.1057} {\bibfield  {journal} {\bibinfo
			{journal} {Rev. Mod. Phys.}\ }\textbf {\bibinfo {volume} {83}},\ \bibinfo
		{pages} {1057} (\bibinfo {year} {2011})}\BibitemShut {NoStop}%
	\bibitem [{\citenamefont {Bernevig}\ and\ \citenamefont
		{Hughes}(2013)}]{Bernevig2013}%
	\BibitemOpen
	\bibfield  {author} {\bibinfo {author} {\bibfnamefont {B.~A.}\ \bibnamefont
			{Bernevig}}\ and\ \bibinfo {author} {\bibfnamefont {T.~L.}\ \bibnamefont
			{Hughes}},\ }\href@noop {} {\emph {\bibinfo {title} {Topological Insulators
				and Topological Superconductors}}}\ (\bibinfo  {publisher} {Princeton
		University Press},\ \bibinfo {address} {Princeton, NJ},\ \bibinfo {year}
	{2013})\BibitemShut {NoStop}%
	\bibitem [{\citenamefont {Berry}(1984)}]{Berry1984}%
	\BibitemOpen
	\bibfield  {author} {\bibinfo {author} {\bibfnamefont {M.~V.}\ \bibnamefont
			{Berry}},\ }\href {\doibase 10.1098/rspa.1984.0023} {\bibfield  {journal}
		{\bibinfo  {journal} {Proc. R. Soc. London, Ser. A}\ }\textbf {\bibinfo
			{volume} {392}},\ \bibinfo {pages} {45} (\bibinfo {year} {1984})}\BibitemShut
	{NoStop}%
	\bibitem [{\citenamefont {Martin}\ \emph {et~al.}(2017)\citenamefont {Martin},
		\citenamefont {Refael},\ and\ \citenamefont {Halperin}}]{Martin_PRX17}%
	\BibitemOpen
	\bibfield  {author} {\bibinfo {author} {\bibfnamefont {I.}~\bibnamefont
			{Martin}}, \bibinfo {author} {\bibfnamefont {G.}~\bibnamefont {Refael}}, \
		and\ \bibinfo {author} {\bibfnamefont {B.}~\bibnamefont {Halperin}},\ }\href
	{\doibase 10.1103/PhysRevX.7.041008} {\bibfield  {journal} {\bibinfo
			{journal} {Phys. Rev. X}\ }\textbf {\bibinfo {volume} {7}},\ \bibinfo {pages}
		{041008} (\bibinfo {year} {2017})}\BibitemShut {NoStop}%
	\bibitem [{\citenamefont {Gaudin}(1976)}]{Gaudin_1976}%
	\BibitemOpen
	\bibfield  {author} {\bibinfo {author} {\bibfnamefont {M.}~\bibnamefont
			{Gaudin}},\ }\href {\doibase 10.1051/jphys:0197600370100108700} {\bibfield
		{journal} {\bibinfo  {journal} {J. Phys. France}\ }\textbf {\bibinfo {volume}
			{37}},\ \bibinfo {pages} {1087} (\bibinfo {year} {1976})}\BibitemShut
	{NoStop}%
	\bibitem [{\citenamefont {Dukelsky}\ \emph {et~al.}(2004)\citenamefont
		{Dukelsky}, \citenamefont {Pittel},\ and\ \citenamefont
		{Sierra}}]{Dukelsky_2004}%
	\BibitemOpen
	\bibfield  {author} {\bibinfo {author} {\bibfnamefont {J.}~\bibnamefont
			{Dukelsky}}, \bibinfo {author} {\bibfnamefont {S.}~\bibnamefont {Pittel}}, \
		and\ \bibinfo {author} {\bibfnamefont {G.}~\bibnamefont {Sierra}},\ }\href
	{\doibase 10.1103/RevModPhys.76.643} {\bibfield  {journal} {\bibinfo
			{journal} {Rev. Mod. Phys.}\ }\textbf {\bibinfo {volume} {76}},\ \bibinfo
		{pages} {643} (\bibinfo {year} {2004})}\BibitemShut {NoStop}%
	\bibitem [{\citenamefont {Claeys}\ \emph {et~al.}(2019)\citenamefont {Claeys},
		\citenamefont {Dimo}, \citenamefont {Baerdemacker},\ and\ \citenamefont
		{Faribault}}]{Claeys_2019}%
	\BibitemOpen
	\bibfield  {author} {\bibinfo {author} {\bibfnamefont {P.~W.}\ \bibnamefont
			{Claeys}}, \bibinfo {author} {\bibfnamefont {C.}~\bibnamefont {Dimo}},
		\bibinfo {author} {\bibfnamefont {S.~D.}\ \bibnamefont {Baerdemacker}}, \
		and\ \bibinfo {author} {\bibfnamefont {A.}~\bibnamefont {Faribault}},\ }\href
	{\doibase 10.1088/1751-8121/aafe9b} {\bibfield  {journal} {\bibinfo
			{journal} {J. Phys. A}\ }\textbf {\bibinfo {volume} {52}},\ \bibinfo {pages}
		{08LT01} (\bibinfo {year} {2019})}\BibitemShut {NoStop}%
	\bibitem [{\citenamefont {Rigolin}\ \emph {et~al.}(2008)\citenamefont
		{Rigolin}, \citenamefont {Ortiz},\ and\ \citenamefont
		{Ponce}}]{Rigolin_2008}%
	\BibitemOpen
	\bibfield  {author} {\bibinfo {author} {\bibfnamefont {G.}~\bibnamefont
			{Rigolin}}, \bibinfo {author} {\bibfnamefont {G.}~\bibnamefont {Ortiz}}, \
		and\ \bibinfo {author} {\bibfnamefont {V.~H.}\ \bibnamefont {Ponce}},\ }\href
	{\doibase 10.1103/PhysRevA.78.052508} {\bibfield  {journal} {\bibinfo
			{journal} {Phys. Rev. A}\ }\textbf {\bibinfo {volume} {78}},\ \bibinfo
		{pages} {052508} (\bibinfo {year} {2008})}\BibitemShut {NoStop}%
	\bibitem [{\citenamefont {Weinberg}\ \emph {et~al.}(2017)\citenamefont
		{Weinberg}, \citenamefont {Bukov}, \citenamefont {D’Alessio}, \citenamefont
		{Polkovnikov}, \citenamefont {Vajna},\ and\ \citenamefont
		{Kolodrubetz}}]{Weinberg_2017}%
	\BibitemOpen
	\bibfield  {author} {\bibinfo {author} {\bibfnamefont {P.}~\bibnamefont
			{Weinberg}}, \bibinfo {author} {\bibfnamefont {M.}~\bibnamefont {Bukov}},
		\bibinfo {author} {\bibfnamefont {L.}~\bibnamefont {D’Alessio}}, \bibinfo
		{author} {\bibfnamefont {A.}~\bibnamefont {Polkovnikov}}, \bibinfo {author}
		{\bibfnamefont {S.}~\bibnamefont {Vajna}}, \ and\ \bibinfo {author}
		{\bibfnamefont {M.}~\bibnamefont {Kolodrubetz}},\ }\href
	{https://doi.org/10.1016/j.physrep.2017.05.003} {\bibfield  {journal}
		{\bibinfo  {journal} {Phys. Rep.}\ }\textbf {\bibinfo {volume} {688}},\
		\bibinfo {pages} {1} (\bibinfo {year} {2017})}\BibitemShut {NoStop}%
	\bibitem [{Note1()}]{Note1}%
	\BibitemOpen
	\bibinfo {note} {In lateral quantum dots or NV centers, the Heisenberg
		interaction to the surrounding spins is not necessarily isotropic, as assumed
		for simplicity in Eq.~\protect \textup {\hbox {\mathsurround \z@ \protect
				\normalfont (\ignorespaces \ref {eq:CentralSpinModel}\unskip \@@italiccorr
				)}}. However, as our main statements about the topological burning glass
		effect reflect fundamental concepts that do not depend on the details of the
		model, our results generally apply to anisotropic interactions as
		well.}\BibitemShut {Stop}%
	\bibitem [{\citenamefont {Hanson}\ \emph {et~al.}(2007)\citenamefont {Hanson},
		\citenamefont {Kouwenhoven}, \citenamefont {Petta}, \citenamefont {Tarucha},\
		and\ \citenamefont {Vandersypen}}]{Hanson_2007}%
	\BibitemOpen
	\bibfield  {author} {\bibinfo {author} {\bibfnamefont {R.}~\bibnamefont
			{Hanson}}, \bibinfo {author} {\bibfnamefont {L.~P.}\ \bibnamefont
			{Kouwenhoven}}, \bibinfo {author} {\bibfnamefont {J.~R.}\ \bibnamefont
			{Petta}}, \bibinfo {author} {\bibfnamefont {S.}~\bibnamefont {Tarucha}}, \
		and\ \bibinfo {author} {\bibfnamefont {L.~M.~K.}\ \bibnamefont
			{Vandersypen}},\ }\href {\doibase 10.1103/RevModPhys.79.1217} {\bibfield
		{journal} {\bibinfo  {journal} {Rev. Mod. Phys.}\ }\textbf {\bibinfo {volume}
			{79}},\ \bibinfo {pages} {1217} (\bibinfo {year} {2007})}\BibitemShut
	{NoStop}%
	\bibitem [{\citenamefont {Doherty}\ \emph {et~al.}(2013)\citenamefont
		{Doherty}, \citenamefont {Manson}, \citenamefont {Delaney}, \citenamefont
		{Jelezko}, \citenamefont {Wrachtrup},\ and\ \citenamefont
		{Hollenberg}}]{Doherty_2013}%
	\BibitemOpen
	\bibfield  {author} {\bibinfo {author} {\bibfnamefont {M.~W.}\ \bibnamefont
			{Doherty}}, \bibinfo {author} {\bibfnamefont {N.~B.}\ \bibnamefont {Manson}},
		\bibinfo {author} {\bibfnamefont {P.}~\bibnamefont {Delaney}}, \bibinfo
		{author} {\bibfnamefont {F.}~\bibnamefont {Jelezko}}, \bibinfo {author}
		{\bibfnamefont {J.}~\bibnamefont {Wrachtrup}}, \ and\ \bibinfo {author}
		{\bibfnamefont {L.~C.}\ \bibnamefont {Hollenberg}},\ }\href {\doibase
		https://doi.org/10.1016/j.physrep.2013.02.001} {\bibfield  {journal}
		{\bibinfo  {journal} {Phys. Rep.}\ }\textbf {\bibinfo {volume} {528}},\
		\bibinfo {pages} {1} (\bibinfo {year} {2013})}\BibitemShut {NoStop}%
	\bibitem [{\citenamefont {Chatterjee}\ \emph {et~al.}(2021)\citenamefont
		{Chatterjee}, \citenamefont {Stevenson}, \citenamefont {De~Franceschi},
		\citenamefont {Morello}, \citenamefont {de~Leon},\ and\ \citenamefont
		{Kuemmeth}}]{Chatterjee_2021}%
	\BibitemOpen
	\bibfield  {author} {\bibinfo {author} {\bibfnamefont {A.}~\bibnamefont
			{Chatterjee}}, \bibinfo {author} {\bibfnamefont {P.}~\bibnamefont
			{Stevenson}}, \bibinfo {author} {\bibfnamefont {S.}~\bibnamefont
			{De~Franceschi}}, \bibinfo {author} {\bibfnamefont {A.}~\bibnamefont
			{Morello}}, \bibinfo {author} {\bibfnamefont {N.~P.}\ \bibnamefont
			{de~Leon}}, \ and\ \bibinfo {author} {\bibfnamefont {F.}~\bibnamefont
			{Kuemmeth}},\ }\href {\doibase 10.1038/s42254-021-00283-9} {\bibfield
		{journal} {\bibinfo  {journal} {Nat. Rev. Phys.}\ }\textbf {\bibinfo {volume}
			{3}},\ \bibinfo {pages} {157} (\bibinfo {year} {2021})}\BibitemShut {NoStop}%
	\bibitem [{\citenamefont {Qi}\ \emph {et~al.}(2006)\citenamefont {Qi},
		\citenamefont {Wu},\ and\ \citenamefont {Zhang}}]{Qi2006}%
	\BibitemOpen
	\bibfield  {author} {\bibinfo {author} {\bibfnamefont {X.-L.}\ \bibnamefont
			{Qi}}, \bibinfo {author} {\bibfnamefont {Y.-S.}\ \bibnamefont {Wu}}, \ and\
		\bibinfo {author} {\bibfnamefont {S.-C.}\ \bibnamefont {Zhang}},\ }\href
	{\doibase 10.1103/PhysRevB.74.085308} {\bibfield  {journal} {\bibinfo
			{journal} {Phys. Rev. B}\ }\textbf {\bibinfo {volume} {74}},\ \bibinfo
		{pages} {085308} (\bibinfo {year} {2006})}\BibitemShut {NoStop}%
	\bibitem [{\citenamefont {Kolodrubetz}\ \emph {et~al.}(2018)\citenamefont
		{Kolodrubetz}, \citenamefont {Nathan}, \citenamefont {Gazit}, \citenamefont
		{Morimoto},\ and\ \citenamefont {Moore}}]{Kolodrubetz2018}%
	\BibitemOpen
	\bibfield  {author} {\bibinfo {author} {\bibfnamefont {M.~H.}\ \bibnamefont
			{Kolodrubetz}}, \bibinfo {author} {\bibfnamefont {F.}~\bibnamefont {Nathan}},
		\bibinfo {author} {\bibfnamefont {S.}~\bibnamefont {Gazit}}, \bibinfo
		{author} {\bibfnamefont {T.}~\bibnamefont {Morimoto}}, \ and\ \bibinfo
		{author} {\bibfnamefont {J.~E.}\ \bibnamefont {Moore}},\ }\href
	{https://doi.org/10.1103/PhysRevLett.120.150601} {\bibfield  {journal}
		{\bibinfo  {journal} {Phys. Rev. Lett.}\ }\textbf {\bibinfo {volume} {120}},\
		\bibinfo {pages} {150601} (\bibinfo {year} {2018})}\BibitemShut {NoStop}%
	\bibitem [{\citenamefont {Peng}\ and\ \citenamefont {Refael}(2018)}]{Peng2018}%
	\BibitemOpen
	\bibfield  {author} {\bibinfo {author} {\bibfnamefont {Y.}~\bibnamefont
			{Peng}}\ and\ \bibinfo {author} {\bibfnamefont {G.}~\bibnamefont {Refael}},\
	}\href {\doibase 10.1103/PhysRevB.97.134303} {\bibfield  {journal} {\bibinfo
			{journal} {Phys. Rev. B}\ }\textbf {\bibinfo {volume} {97}},\ \bibinfo
		{pages} {134303} (\bibinfo {year} {2018})}\BibitemShut {NoStop}%
	\bibitem [{\citenamefont {Crowley}\ \emph {et~al.}(2019)\citenamefont
		{Crowley}, \citenamefont {Martin},\ and\ \citenamefont
		{Chandran}}]{Crowley2019}%
	\BibitemOpen
	\bibfield  {author} {\bibinfo {author} {\bibfnamefont {P.~J.~D.}\
			\bibnamefont {Crowley}}, \bibinfo {author} {\bibfnamefont {I.}~\bibnamefont
			{Martin}}, \ and\ \bibinfo {author} {\bibfnamefont {A.}~\bibnamefont
			{Chandran}},\ }\href {\doibase 10.1103/PhysRevB.99.064306} {\bibfield
		{journal} {\bibinfo  {journal} {Phys. Rev. B}\ }\textbf {\bibinfo {volume}
			{99}},\ \bibinfo {pages} {064306} (\bibinfo {year} {2019})}\BibitemShut
	{NoStop}%
	\bibitem [{\citenamefont {Nathan}\ \emph {et~al.}(2019)\citenamefont {Nathan},
		\citenamefont {Martin},\ and\ \citenamefont {Refael}}]{Nathan2019}%
	\BibitemOpen
	\bibfield  {author} {\bibinfo {author} {\bibfnamefont {F.}~\bibnamefont
			{Nathan}}, \bibinfo {author} {\bibfnamefont {I.}~\bibnamefont {Martin}}, \
		and\ \bibinfo {author} {\bibfnamefont {G.}~\bibnamefont {Refael}},\ }\href
	{\doibase 10.1103/PhysRevB.99.094311} {\bibfield  {journal} {\bibinfo
			{journal} {Phys. Rev. B}\ }\textbf {\bibinfo {volume} {99}},\ \bibinfo
		{pages} {094311} (\bibinfo {year} {2019})}\BibitemShut {NoStop}%
	\bibitem [{\citenamefont {K\"orber}\ \emph {et~al.}(2020)\citenamefont
		{K\"orber}, \citenamefont {Privitera}, \citenamefont {Budich},\ and\
		\citenamefont {Trauzettel}}]{Koerber_2020}%
	\BibitemOpen
	\bibfield  {author} {\bibinfo {author} {\bibfnamefont {S.}~\bibnamefont
			{K\"orber}}, \bibinfo {author} {\bibfnamefont {L.}~\bibnamefont {Privitera}},
		\bibinfo {author} {\bibfnamefont {J.~C.}\ \bibnamefont {Budich}}, \ and\
		\bibinfo {author} {\bibfnamefont {B.}~\bibnamefont {Trauzettel}},\ }\href
	{\doibase 10.1103/PhysRevResearch.2.022023} {\bibfield  {journal} {\bibinfo
			{journal} {Phys. Rev. Research}\ }\textbf {\bibinfo {volume} {2}},\ \bibinfo
		{pages} {022023(R)} (\bibinfo {year} {2020})}\BibitemShut {NoStop}%
	\bibitem [{\citenamefont {Crowley}\ \emph {et~al.}(2020)\citenamefont
		{Crowley}, \citenamefont {Martin},\ and\ \citenamefont
		{Chandran}}]{Crowley_2020}%
	\BibitemOpen
	\bibfield  {author} {\bibinfo {author} {\bibfnamefont {P.~J.~D.}\
			\bibnamefont {Crowley}}, \bibinfo {author} {\bibfnamefont {I.}~\bibnamefont
			{Martin}}, \ and\ \bibinfo {author} {\bibfnamefont {A.}~\bibnamefont
			{Chandran}},\ }\href {\doibase 10.1103/PhysRevLett.125.100601} {\bibfield
		{journal} {\bibinfo  {journal} {Phys. Rev. Lett.}\ }\textbf {\bibinfo
			{volume} {125}},\ \bibinfo {pages} {100601} (\bibinfo {year}
		{2020})}\BibitemShut {NoStop}%
	\bibitem [{\citenamefont {Boyers}\ \emph {et~al.}(2020)\citenamefont {Boyers},
		\citenamefont {Crowley}, \citenamefont {Chandran},\ and\ \citenamefont
		{Sushkov}}]{Boyers_2020}%
	\BibitemOpen
	\bibfield  {author} {\bibinfo {author} {\bibfnamefont {E.}~\bibnamefont
			{Boyers}}, \bibinfo {author} {\bibfnamefont {P.~J.~D.}\ \bibnamefont
			{Crowley}}, \bibinfo {author} {\bibfnamefont {A.}~\bibnamefont {Chandran}}, \
		and\ \bibinfo {author} {\bibfnamefont {A.~O.}\ \bibnamefont {Sushkov}},\
	}\href {\doibase 10.1103/PhysRevLett.125.160505} {\bibfield  {journal}
		{\bibinfo  {journal} {Phys. Rev. Lett.}\ }\textbf {\bibinfo {volume} {125}},\
		\bibinfo {pages} {160505} (\bibinfo {year} {2020})}\BibitemShut {NoStop}%
	\bibitem [{\citenamefont {Chen}\ \emph {et~al.}(2020)\citenamefont {Chen},
		\citenamefont {Liu}, \citenamefont {Yu}, \citenamefont {Zhang},\ and\
		\citenamefont {Cai}}]{Chen_2020}%
	\BibitemOpen
	\bibfield  {author} {\bibinfo {author} {\bibfnamefont {Q.}~\bibnamefont
			{Chen}}, \bibinfo {author} {\bibfnamefont {H.}~\bibnamefont {Liu}}, \bibinfo
		{author} {\bibfnamefont {M.}~\bibnamefont {Yu}}, \bibinfo {author}
		{\bibfnamefont {S.}~\bibnamefont {Zhang}}, \ and\ \bibinfo {author}
		{\bibfnamefont {J.}~\bibnamefont {Cai}},\ }\href
	{https://doi.org/10.1103/PhysRevA.102.052606} {\bibfield  {journal} {\bibinfo
			{journal} {Phys. Rev. A}\ }\textbf {\bibinfo {volume} {102}},\ \bibinfo
		{pages} {052606} (\bibinfo {year} {2020})}\BibitemShut {NoStop}%
	\bibitem [{\citenamefont {Nathan}\ \emph {et~al.}(2020)\citenamefont {Nathan},
		\citenamefont {Refael}, \citenamefont {Rudner},\ and\ \citenamefont
		{Martin}}]{Nathan_2020}%
	\BibitemOpen
	\bibfield  {author} {\bibinfo {author} {\bibfnamefont {F.}~\bibnamefont
			{Nathan}}, \bibinfo {author} {\bibfnamefont {G.}~\bibnamefont {Refael}},
		\bibinfo {author} {\bibfnamefont {M.~S.}\ \bibnamefont {Rudner}}, \ and\
		\bibinfo {author} {\bibfnamefont {I.}~\bibnamefont {Martin}},\ }\href
	{\doibase 10.1103/PhysRevResearch.2.043411} {\bibfield  {journal} {\bibinfo
			{journal} {Phys. Rev. Research}\ }\textbf {\bibinfo {volume} {2}},\ \bibinfo
		{pages} {043411} (\bibinfo {year} {2020})}\BibitemShut {NoStop}%
	\bibitem [{\citenamefont {Long}\ \emph {et~al.}(2021)\citenamefont {Long},
		\citenamefont {Crowley},\ and\ \citenamefont {Chandran}}]{Long_2021}%
	\BibitemOpen
	\bibfield  {author} {\bibinfo {author} {\bibfnamefont {D.~M.}\ \bibnamefont
			{Long}}, \bibinfo {author} {\bibfnamefont {P.~J.~D.}\ \bibnamefont
			{Crowley}}, \ and\ \bibinfo {author} {\bibfnamefont {A.}~\bibnamefont
			{Chandran}},\ }\href {\doibase 10.1103/PhysRevLett.126.106805} {\bibfield
		{journal} {\bibinfo  {journal} {Phys. Rev. Lett.}\ }\textbf {\bibinfo
			{volume} {126}},\ \bibinfo {pages} {106805} (\bibinfo {year}
		{2021})}\BibitemShut {NoStop}%
	\bibitem [{\citenamefont {Nathan}\ \emph {et~al.}(2021)\citenamefont {Nathan},
		\citenamefont {Ge}, \citenamefont {Gazit}, \citenamefont {Rudner},\ and\
		\citenamefont {Kolodrubetz}}]{Nathan_2021}%
	\BibitemOpen
	\bibfield  {author} {\bibinfo {author} {\bibfnamefont {F.}~\bibnamefont
			{Nathan}}, \bibinfo {author} {\bibfnamefont {R.}~\bibnamefont {Ge}}, \bibinfo
		{author} {\bibfnamefont {S.}~\bibnamefont {Gazit}}, \bibinfo {author}
		{\bibfnamefont {M.}~\bibnamefont {Rudner}}, \ and\ \bibinfo {author}
		{\bibfnamefont {M.}~\bibnamefont {Kolodrubetz}},\ }\href
	{https://doi.org/10.1103/PhysRevLett.127.166804} {\bibfield  {journal}
		{\bibinfo  {journal} {Phys. Rev. Lett.}\ }\textbf {\bibinfo {volume} {127}},\
		\bibinfo {pages} {166804} (\bibinfo {year} {2021})}\BibitemShut {NoStop}%
	\bibitem [{\citenamefont {Psaroudaki}\ and\ \citenamefont
		{Refael}(2021)}]{Psaroudaki_2021}%
	\BibitemOpen
	\bibfield  {author} {\bibinfo {author} {\bibfnamefont {C.}~\bibnamefont
			{Psaroudaki}}\ and\ \bibinfo {author} {\bibfnamefont {G.}~\bibnamefont
			{Refael}},\ }\href {https://doi.org/10.1016/j.aop.2021.168553} {\bibfield
		{journal} {\bibinfo  {journal} {Annals of Physics}\ }\textbf {\bibinfo
			{volume} {435}},\ \bibinfo {pages} {168553} (\bibinfo {year}
		{2021})}\BibitemShut {NoStop}%
	\bibitem [{\citenamefont {Qi}\ \emph {et~al.}(2021)\citenamefont {Qi},
		\citenamefont {Refael},\ and\ \citenamefont {Peng}}]{Qi_2021}%
	\BibitemOpen
	\bibfield  {author} {\bibinfo {author} {\bibfnamefont {Z.}~\bibnamefont
			{Qi}}, \bibinfo {author} {\bibfnamefont {G.}~\bibnamefont {Refael}}, \ and\
		\bibinfo {author} {\bibfnamefont {Y.}~\bibnamefont {Peng}},\ }\href
	{https://doi.org/10.1103/PhysRevB.104.224301} {\bibfield  {journal} {\bibinfo
			{journal} {Phys. Rev. B}\ }\textbf {\bibinfo {volume} {104}},\ \bibinfo
		{pages} {224301} (\bibinfo {year} {2021})}\BibitemShut {NoStop}%
	\bibitem [{\citenamefont {Schwennicke}\ and\ \citenamefont
		{Yuen-Zhou}(2022)}]{Schwennicke_2022}%
	\BibitemOpen
	\bibfield  {author} {\bibinfo {author} {\bibfnamefont {K.}~\bibnamefont
			{Schwennicke}}\ and\ \bibinfo {author} {\bibfnamefont {J.}~\bibnamefont
			{Yuen-Zhou}},\ }\href {\doibase 10.1021/acs.jpclett.1c04161} {\bibfield
		{journal} {\bibinfo  {journal} {J. Phys. Chem. Lett.}\ }\textbf {\bibinfo
			{volume} {13}},\ \bibinfo {pages} {2434} (\bibinfo {year}
		{2022})}\BibitemShut {NoStop}%
	\bibitem [{\citenamefont {Gritsev}\ and\ \citenamefont
		{Polkovnikov}(2012)}]{Gritsev_2012}%
	\BibitemOpen
	\bibfield  {author} {\bibinfo {author} {\bibfnamefont {V.}~\bibnamefont
			{Gritsev}}\ and\ \bibinfo {author} {\bibfnamefont {A.}~\bibnamefont
			{Polkovnikov}},\ }\href {\doibase 10.1073/pnas.1116693109} {\bibfield
		{journal} {\bibinfo  {journal} {Proceedings of the National Academy of
				Sciences}\ }\textbf {\bibinfo {volume} {109}},\ \bibinfo {pages} {6457}
		(\bibinfo {year} {2012})}\BibitemShut {NoStop}%
	\bibitem [{\citenamefont {Bloch}(1929)}]{Bloch_1929}%
	\BibitemOpen
	\bibfield  {author} {\bibinfo {author} {\bibfnamefont {F.}~\bibnamefont
			{Bloch}},\ }\href {https://doi.org/10.1007/BF01339455} {\bibfield  {journal}
		{\bibinfo  {journal} {Zeitschrift für Physik}\ }\textbf {\bibinfo {volume}
			{52}},\ \bibinfo {pages} {555} (\bibinfo {year} {1929})}\BibitemShut
	{NoStop}%
	\bibitem [{Note2()}]{Note2}%
	\BibitemOpen
	\bibinfo {note} {Note that Eq.~\protect \textup {\hbox {\mathsurround \z@
				\protect \normalfont (\ignorespaces \ref {eq:ExpectationValue}\unskip
				\@@italiccorr )}} also holds for the exact dynamics of the system, involving
		the higher orders in the perturbative expansion of Eq.~\protect \textup
		{\hbox {\mathsurround \z@ \protect \normalfont (\ignorespaces \ref
				{eq:AdiabaticPerturbationTheory}\unskip \@@italiccorr )}}.}\BibitemShut
	{Stop}%
	\bibitem [{Note3()}]{Note3}%
	\BibitemOpen
	\bibinfo {note} {Results for antiferromagnetic coupling strength $A<0$ are
		presented in the Appendix~\ref
		{sec:AntiferromagneticInteraction}}\BibitemShut {NoStop}%
	\bibitem [{\citenamefont {Schroer}\ \emph {et~al.}(2014)\citenamefont
		{Schroer}, \citenamefont {Kolodrubetz}, \citenamefont {Kindel}, \citenamefont
		{Sandberg}, \citenamefont {Gao}, \citenamefont {Vissers}, \citenamefont
		{Pappas}, \citenamefont {Polkovnikov},\ and\ \citenamefont
		{Lehnert}}]{Schroer_2014}%
	\BibitemOpen
	\bibfield  {author} {\bibinfo {author} {\bibfnamefont {M.~D.}\ \bibnamefont
			{Schroer}}, \bibinfo {author} {\bibfnamefont {M.~H.}\ \bibnamefont
			{Kolodrubetz}}, \bibinfo {author} {\bibfnamefont {W.~F.}\ \bibnamefont
			{Kindel}}, \bibinfo {author} {\bibfnamefont {M.}~\bibnamefont {Sandberg}},
		\bibinfo {author} {\bibfnamefont {J.}~\bibnamefont {Gao}}, \bibinfo {author}
		{\bibfnamefont {M.~R.}\ \bibnamefont {Vissers}}, \bibinfo {author}
		{\bibfnamefont {D.~P.}\ \bibnamefont {Pappas}}, \bibinfo {author}
		{\bibfnamefont {A.}~\bibnamefont {Polkovnikov}}, \ and\ \bibinfo {author}
		{\bibfnamefont {K.~W.}\ \bibnamefont {Lehnert}},\ }\href
	{https://doi.org/10.1103/PhysRevLett.113.050402} {\bibfield  {journal}
		{\bibinfo  {journal} {Phys. Rev. Lett.}\ }\textbf {\bibinfo {volume} {113}},\
		\bibinfo {pages} {050402} (\bibinfo {year} {2014})}\BibitemShut {NoStop}%
	\bibitem [{\citenamefont {Roushan}\ \emph {et~al.}(2014)\citenamefont
		{Roushan}, \citenamefont {Neill}, \citenamefont {Chen}, \citenamefont
		{Kolodrubetz}, \citenamefont {Quintana}, \citenamefont {Leung}, \citenamefont
		{Fang}, \citenamefont {Barends}, \citenamefont {Campbell}, \citenamefont
		{Chen}, \citenamefont {Chiaro}, \citenamefont {Dunsworth}, \citenamefont
		{Jeffrey}, \citenamefont {Kelly}, \citenamefont {Megrant}, \citenamefont
		{Mutus}, \citenamefont {O'Malley}, \citenamefont {Sank}, \citenamefont
		{Vainsencher}, \citenamefont {Wenner}, \citenamefont {White}, \citenamefont
		{Polkovnikov}, \citenamefont {Cleland},\ and\ \citenamefont
		{Martinis}}]{Roushan2014}%
	\BibitemOpen
	\bibfield  {author} {\bibinfo {author} {\bibfnamefont {P.}~\bibnamefont
			{Roushan}}, \bibinfo {author} {\bibfnamefont {C.}~\bibnamefont {Neill}},
		\bibinfo {author} {\bibfnamefont {Y.}~\bibnamefont {Chen}}, \bibinfo {author}
		{\bibfnamefont {M.}~\bibnamefont {Kolodrubetz}}, \bibinfo {author}
		{\bibfnamefont {C.}~\bibnamefont {Quintana}}, \bibinfo {author}
		{\bibfnamefont {N.}~\bibnamefont {Leung}}, \bibinfo {author} {\bibfnamefont
			{M.}~\bibnamefont {Fang}}, \bibinfo {author} {\bibfnamefont {R.}~\bibnamefont
			{Barends}}, \bibinfo {author} {\bibfnamefont {B.}~\bibnamefont {Campbell}},
		\bibinfo {author} {\bibfnamefont {Z.}~\bibnamefont {Chen}}, \bibinfo {author}
		{\bibfnamefont {B.}~\bibnamefont {Chiaro}}, \bibinfo {author} {\bibfnamefont
			{A.}~\bibnamefont {Dunsworth}}, \bibinfo {author} {\bibfnamefont
			{E.}~\bibnamefont {Jeffrey}}, \bibinfo {author} {\bibfnamefont
			{J.}~\bibnamefont {Kelly}}, \bibinfo {author} {\bibfnamefont
			{A.}~\bibnamefont {Megrant}}, \bibinfo {author} {\bibfnamefont
			{J.}~\bibnamefont {Mutus}}, \bibinfo {author} {\bibfnamefont {P.~J.~J.}\
			\bibnamefont {O'Malley}}, \bibinfo {author} {\bibfnamefont {D.}~\bibnamefont
			{Sank}}, \bibinfo {author} {\bibfnamefont {A.}~\bibnamefont {Vainsencher}},
		\bibinfo {author} {\bibfnamefont {J.}~\bibnamefont {Wenner}}, \bibinfo
		{author} {\bibfnamefont {T.}~\bibnamefont {White}}, \bibinfo {author}
		{\bibfnamefont {A.}~\bibnamefont {Polkovnikov}}, \bibinfo {author}
		{\bibfnamefont {A.~N.}\ \bibnamefont {Cleland}}, \ and\ \bibinfo {author}
		{\bibfnamefont {J.~M.}\ \bibnamefont {Martinis}},\ }\href
	{https://doi.org/10.1038/nature13891} {\bibfield  {journal} {\bibinfo
			{journal} {Nature}\ }\textbf {\bibinfo {volume} {515}},\ \bibinfo {pages}
		{241} (\bibinfo {year} {2014})}\BibitemShut {NoStop}%
	\bibitem [{\citenamefont {Malz}\ and\ \citenamefont {Smith}(2021)}]{Malz_2021}%
	\BibitemOpen
	\bibfield  {author} {\bibinfo {author} {\bibfnamefont {D.}~\bibnamefont
			{Malz}}\ and\ \bibinfo {author} {\bibfnamefont {A.}~\bibnamefont {Smith}},\
	}\href {\doibase 10.1103/PhysRevLett.126.163602} {\bibfield  {journal}
		{\bibinfo  {journal} {Phys. Rev. Lett.}\ }\textbf {\bibinfo {volume} {126}},\
		\bibinfo {pages} {163602} (\bibinfo {year} {2021})}\BibitemShut {NoStop}%
	\bibitem [{Note4()}]{Note4}%
	\BibitemOpen
	\bibinfo {note} {For a time $\lambda \protect \tmspace +\thinmuskip
		{.1667em}T=5\cdot 10^5$, the entire synthetic 2D Brillouin zone (BZ) is
		sufficiently sampled for our choice of parameters.}\BibitemShut {Stop}%
	\bibitem [{Note5()}]{Note5}%
	\BibitemOpen
	\bibinfo {note} {We obtain the gradient by computing $P^{12}=\protect \frac
		{2}{T}\protect \tmspace +\thinmuskip {.1667em}(\protect \mathaccentV
		{bar}016{W}_{1}^{T}-\protect \mathaccentV {bar}016{W}_{1}^{T/2})$ with the
		time-averaged energy transfers $\protect \mathaccentV
		{bar}016{W}_{1}^{T/2}=\protect \frac {2}{T}\protect \tmspace +\thinmuskip
		{.1667em}\DOTSI \intop \ilimits@ _{0}^{T/2}\protect \tmspace +\thinmuskip
		{.1667em}\protect \mathrm {d}t\protect \tmspace +\thinmuskip
		{.1667em}W_{1}(t)$ and $\protect \mathaccentV {bar}016{W}_{1}^{T}=\protect
		\frac {2}{T}\protect \tmspace +\thinmuskip {.1667em}\DOTSI \intop \ilimits@
		_{T/2}^{T}\protect \tmspace +\thinmuskip {.1667em}\protect \mathrm
		{d}t\protect \tmspace +\thinmuskip {.1667em}W_{1}(t)$.}\BibitemShut {Stop}%
	\bibitem [{\citenamefont {Nathan}\ \emph {et~al.}(2022)\citenamefont {Nathan},
		\citenamefont {Martin},\ and\ \citenamefont {Refael}}]{Nathan_2022}%
	\BibitemOpen
	\bibfield  {author} {\bibinfo {author} {\bibfnamefont {F.}~\bibnamefont
			{Nathan}}, \bibinfo {author} {\bibfnamefont {I.}~\bibnamefont {Martin}}, \
		and\ \bibinfo {author} {\bibfnamefont {G.}~\bibnamefont {Refael}},\ }\href
	{https://arxiv.org/abs/2201.07804} {\bibfield  {journal} {\bibinfo  {journal}
			{arXiv:2201.07804}\ } (\bibinfo {year} {2022})}\BibitemShut {NoStop}%
\end{thebibliography}
\end{document}